\documentstyle[epsf,aps,multicol]{revtex}
\begin{document}
\title{Structure-Property Correlations in Model Composite Materials}
\author{A. P. Roberts and M. A. Knackstedt}
\address{
Department of Applied Mathematics,
Research School of Physical Sciences,
Australian National University, \\
Canberra,
Australian Capital Territory,
0200 Australia }
\date{9 April 1996}
\maketitle
\begin{abstract}
We investigate the effective properties
(conductivity, diffusivity and elastic
moduli) of model random composite media derived
from Gaussian random fields
and overlapping hollow spheres.
The morphologies generated in the models 
exhibit low percolation thresholds and give a realistic
representation of
the complex microstructure observed
in many classes of composites.
The statistical correlation functions of the
models are derived
and used to evaluate rigorous bounds on each property. 
Simulation of the effective conductivity is used to
demonstrate the
applicability of the bounds.
The key morphological features which
effect composite properties are discussed.
\end{abstract}
\pacs{05.40.+j, 61.43.-j, 72.15.Eb, 62.20.Dc}

\begin{multicols}{2}

\section{Introduction}
\label{introch6}
\begin{picture}(0,0)
\put (35,-446){ \parbox{16.0cm}{\small \raggedleft %
{\em Appeared in} PHYSICAL REVIEW E, VOL.\ 54, %
PAGES 2313$\;$-$\;$2328, SEPTEMBER 1996}}
\end{picture}
The prediction of effective properties of heterogeneous systems such
as porous media and two phase composites is of considerable
interest~\cite{TorqRev91,Isichenko92,Sahimi93}. 
Understanding the inter-relationships between rock properties
and their expression in geophysical and petrophysical data is
necessary for enhanced characterisation of underground reservoirs.
This understanding is crucial to the economics of oil and gas
recovery, geothermal energy extraction and groundwater pollution
abatement.
Manufactured
composites such as foamed solids~\cite{Gibson88} and
polymer blends~\cite{Gubbels94} often exhibit a complex
microstructure.
To optimize the properties of these systems it is necessary to understand how
morphology influences effective properties.
In general, the
difficulty of accounting for microstructure has
made exact prediction impossible in all but the simplest of cases.

\begin{figure}[b!]

\hskip 1mm
\centering \epsfxsize=14cm\epsfbox{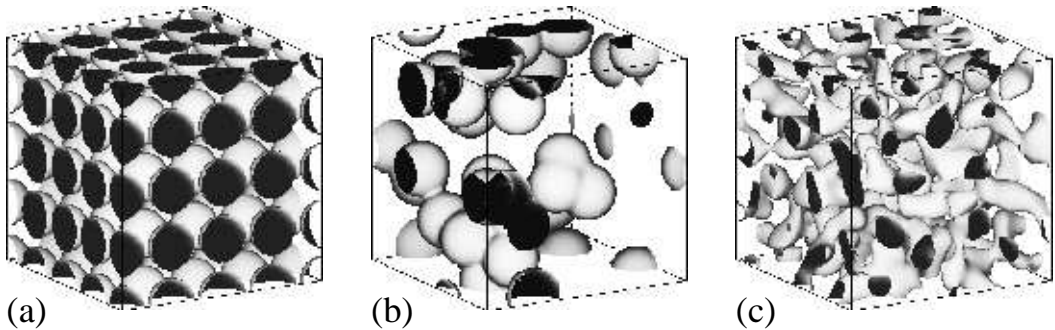}

\vspace{2mm}

\caption{\sl Models of composite microstructure. (a) Periodic models
- regular array of spheres; (b) Particulate models - identical
overlapping spheres (IOS) and;
(c) Gaussian random field (GRF) models - single
cut variant. 
None of the models can mimic the microstructure observed in
percolative low volume fraction materials (e.g.\ polystyrene foam - 
Fig.~\ref{polysty}). 
\label{3Dsimple}
\label{IIIb_0p_2}}
\end{figure}

On the other hand, considerable progress has been made in
the derivation of rigorous bounds on a host
of properties~\cite{TorqRev91,TorqRev94}.
For example, relatively accurate bounds have been derived for the elastic
moduli and conductivity of isotropic two-phase 
composites~\cite{Beran65a,Beran65b,McCoy,Milton81b,Milton82b}.
To evaluate these bounds for a given system it is necessary
to know the 3-point statistical correlation function~\cite{Milton81a}.
Due to the difficulty of measuring this
information~\cite{CorsonI,Adler92,Berryman86}, a number
of model media have been proposed for which the functions
can be explicitly evaluated. These include: cellular~\cite{Miller69},
particulate~\cite{TorqRev91} and periodic~\cite{McPhedran81} materials
(eg.\ Figs.~\ref{3Dsimple}(a)\&(b)).
The principal problem with these models 
is that they employ over-simplified representations of
the inclusion (or pore) structure observed in many natural and manufactured
composite materials.

Recently we derived the properties of 
a model of amorphous materials~\cite{Roberts95a}
(e.g.\ Fig.~\ref{3Dsimple}(c)) based on
level-cut Gaussian random fields~\cite{Berk87} (GRF).
Although the GRF model is applicable to many classes of
non-particulate composite materials, it cannot account
for materials which remain percolative at very low volume fractions.

\begin{figure}[t!]

\hskip 0mm
{\samepage\columnwidth20.5pc
\centering \epsfxsize=6.0cm\epsfbox{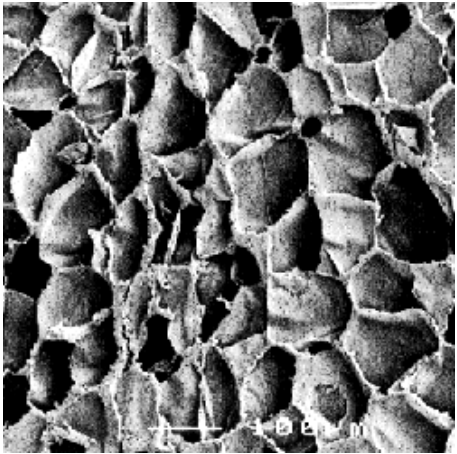}
\vspace{4mm}
\caption{\sl Morphology of polystyrene foam.
\label{polysty}}
}
\end{figure}

Porous rocks~\cite{Sahimi93,Wong84}, polymer blends~\cite{Gubbels94}, 
solid foams~\cite{Gibson88} and membranes provide
examples of systems where a single phase remains connected down to
low volume fractions. The percolation threshold of a system is
only one factor which determines its effective properties:
The shape of the pores/inclusions should also be
considered~\cite{Bernabe91,Saeger91,Roberts85}.
Polystyrene foam, an example of a highly-porous
material, is shown in Fig.~\ref{polysty}.

\end{multicols}

\begin{multicols}{2}

\vspace{2mm}

\noindent
The complex solid phase has a `sheet-like' character
quite different to that found in cellular, particulate~\cite{TorqRev91}
and single level cut GRF~\cite{Adler92,Roberts95a} models 
(Fig.~\ref{3Dsimple}).
It is clear that current models of composite microstructure cannot
account for the percolative and morphological characteristics
observed in porous rocks, solid foams, membranes
and polymer blends.

In this paper we describe models which do give
a realistic representation of the microstructure observed in
many classes of composite materials, and which remain percolative
at very low volume fractions.
Variational bounds and computer simulation
are used to estimate the influence of morphology on
diffusive-transport and elastic properties.
The first model is an extension of the Gaussian
random field (GRF) model considered in a
previous paper~\cite{Roberts95a}.
In this case the interface between the
composite phases is defined by two (rather than one)
level cut of a GRF~\cite{Berk87,Quiblier84,Berk91,Teubner91}.
The freedom in choosing the position of the cuts (for a given volume
fraction), and the spectra of the model, allows a rich variety
of morphologies to be modelled. By qualitatively
comparing these morphologies to those observed in physical systems
the models can be associated with classes of physical composites. 

A second highly porous model can be obtained by generalizing
the well-known IOS model~\cite{TorqRev91} to include the case
of arbitrarily thin hollow spheres. This model is applicable to
a class of ceramics and foams fabricated from hollow
spheres: a composite which possesses excellent uniformity and
properties~\cite{Green85}. 

To study the properties of these media we evaluate bounds on the effective
conductivity and elastic moduli. The key microstructure
parameters ($\zeta_1$ \& $\eta_1$) which occur in the derivation 
of the bounds~\cite{Milton81a} are tabulated along with 
illustrations of the model morphologies.
In addition we use a finite difference scheme to directly
simulate the effective conductivity.  This allows us to comment on
the applicability of the bounds, and on their use for predictive purposes.

The paper is organized as follows.
In Section \ref{correl_ab} we derive the 3-point correlation function
for the 2-level cut Gaussian random field.
In Section \ref{iosa} analogous results are
derived for the `identical overlapping spherical annuli' (IOSA) or
`hollow sphere' model. In Sections \ref{zeta_ab} and \ref{simuls_ab}
the microstructure parameters are calculated, and computer simulations of
the effective conductivity are compared with the resultant bounds. 
In section \ref{influence} we discuss the influence of morphology on
the transport and mechanical properties of composites.
\section{The 2-level cut GRF model}
\label{correl_ab}
As in \cite{Roberts95a} we take $y({\mathbf r})$ as an isotropic
Gaussian random field with a given field-field correlation function
$\langle y({\mathbf r}_i)y({\mathbf r}_j)\rangle =g_{\mbox{\tiny$K$}}(r_{ij})$. Here
$r_{ij}=|{\mathbf r}_i-{\mathbf r}_j|$ and for convenience
we denote $g_{\mbox{\tiny$K$}}(r_{ij})$
by $g_{ij}$, or simply $g$ if no ambiguity arises.
Following Berk \cite{Berk87} it is
possible to define a composite medium with phase 1 the region in space
where $\alpha \leq y({\mathbf r}) \leq \beta$. The remaining region is phase
2. In the limit $\beta\to\infty$ the 1-level cut GRF considered
in Refs.~\cite{Roberts95a,Teubner91} is recovered.
The $n$-point correlation function is given by
\begin{equation}\label{defnpn6}
p_n({\mathbf r}_1,{\mathbf r}_2,\dots,{\mathbf r}_n)=
\left\langle \prod_{i=1}^{n} \left[ H(y_i-\alpha)-H(y_i-\beta) \right]
\right\rangle, \end{equation}
where $H(y)$ is the Heaviside function and $y_i=y({\mathbf r}_i)$.

The microstructure of the material is fully determined by specifying $\alpha$,
$\beta$ and $g_{\mbox{\tiny$K$}}(r)$. The latter quantity is
related to the spectral density of the field $\rho_{\mbox{\tiny$K$}}(k)$
by a clipped Fourier transform:
\begin{equation} \label{gK}
g_{\mbox{\tiny$K$}}(r)=\int_0^K 4\pi k^2 \rho_{\mbox{\tiny$K$}}(k)
\frac{ \sin kr}{kr} dk.
\end{equation}
It was shown in \cite{Roberts95a} that few
differences arise amongst the conductivity of the 1-level cut
Gaussian random fields defined with differing spectra. Therefore we employ
two model materials which showed the greatest variation in properties.
In the notation of \cite{Roberts95a} these are Model I:
\begin{eqnarray}
\rho(k)&=&P^{-1}\pi^{-2}\left( (1-\nu^2+k^2)^2+4\nu^2 \right)^{-1} \\
\lim_{K\to\infty} g_{\mbox{\tiny$K$}}(r)&=&e^{-r}\frac{\sin\nu r}{\nu r},
\end{eqnarray}   
where $P$ is a normalization constant chosen to
ensure $g_{\mbox{\tiny$K$}}(0)=1$, and Model III:
\begin{eqnarray}
\rho(k)&=&\frac{3}{4\pi(\mu^3-1)} [ H(\mu)-H(1)] \;\; (\mu > 1) \\
g(r)&=& \frac
{3\left(\sin \mu r-\mu r\cos \mu r-\sin r+r\cos r \right)}
{r^3(\mu^3-1)}.
\end{eqnarray} 
No normalization constant is necessary in this model provided that $K\geq\mu$.
In following sections we employ spectrum I ($\nu=0$,
$K=\infty$), spectrum I ($\nu=0$, $K=8$) and
spectrum III ($\mu=1.5$). In this paper the
physical parameters $\nu$ and $\mu$ are not varied and will no longer be
explicitly stated.

In the notation of Appendix \ref{maxematics} the 1-point correlation
function (or volume fraction) is just
\begin{equation} p=\Lambda_1(\alpha)-\Lambda_1(\beta)=
\frac1{\sqrt{2\pi}}\int_\alpha^\beta e^{-\frac12 t^2} dt \label{ab1pnt}. \end{equation}
The 2-point correlation function for the 2-level cut Gaussian random
field can be defined as
\begin{equation} 
p_2=p^2+\Lambda_2(g,\alpha,\alpha)-2\Lambda_2(g,\alpha,\beta)+
\Lambda_2(g,\beta,\beta)
\end{equation}
where we have used the fact that
$\Lambda_2(g,\alpha,\beta)=\Lambda_2(g,\beta,\alpha)$ and $p_2(g=0)=p^2$.
Now using Eqn.~(\ref{lam2}) leads to~\cite{Berk91}
\begin{eqnarray}
&&\lefteqn{p_2(g)=p^2+\frac{1}{2\pi}\int_0^g
 \frac{dt}{\sqrt{1-t^2}} \times  \left[
\exp\left({-\frac{\alpha^2}{1+t}}\right) \right. }  \nonumber 
\\ && \left. 
-\exp\left({-\frac12 \frac{\alpha^2-2\alpha\beta t+\beta^2}{(1-t^2)}}\right)
+\exp\left({-\frac{\beta^2}{1+t}}\right) \right].
\end{eqnarray}
Similarly the 3-point correlation function is,
\begin{eqnarray} p_3=p^3 &+& \nonumber
 \Lambda_3({\mathbf g},\alpha,\alpha,\alpha) 
-\Lambda_3({\mathbf g},\alpha,\alpha,\beta ) 
-\Lambda_3({\mathbf g},\alpha,\beta ,\alpha) \\
\nonumber
&-& \Lambda_3({\mathbf g},\beta ,\alpha,\alpha) 
+\Lambda_3({\mathbf g},\alpha,\beta ,\beta ) 
+\Lambda_3({\mathbf g},\beta ,\alpha,\beta ) \\
&+&\Lambda_3({\mathbf g},\beta ,\beta ,\alpha) 
 - \Lambda_3({\mathbf g},\beta ,\beta ,\beta ) 
\end{eqnarray}
where $\Lambda_3$ is given in Eq.\ \ref{lam3} and ${\mathbf g}=
(g_{12},g_{13},g_{23})$.
We could find no symmetries in these terms to allow
analytical or computational simplification of the results.

For our purposes it is necessary to choose $\alpha$ and $\beta$ for a given
value of the volume fraction $p$. There are many ways that this
can be done.  An obvious method is to require that an
equivalent fraction of phase 1 lies on either side of a
particular level cut $y({\mathbf r})=\gamma$.
We classify these `symmetric' models by the parameter
\begin{equation}\label{class_s}
s=\frac1{\sqrt{2\pi}}\int_\gamma^\infty e^{-\frac12 t^2} dt \end{equation}
so that $s\in[0,1]$.
For a given volume fraction $p$, $\alpha$ and $\beta$ are defined through the
relations
\begin{equation}
\frac p2 = \frac1{\sqrt{2\pi}}\int_\alpha^\gamma e^{-\frac12 t^2} dt =
\frac1{\sqrt{2\pi}}\int_\gamma^\beta e^{-\frac12 t^2} dt
\end{equation}
Materials defined in this manner are denoted by, for example, III(s=.5).
This indicates that the spectrum of model III 
is employed and that $s=0.5$ (corresponding to the case $\alpha=-\beta$).
For comparison with the 1-level cut case discussed in \cite{Roberts95a}
it is also useful to define a 2-level cut GRF which reduces to
the former model in a particular limit.
This is done by fixing $\beta$ and varying
$\alpha$ such that a given volume fraction is achieved.
These `base'-level models are specified by the value
\begin{equation} b=\frac1{\sqrt{2\pi}}\int_\beta^\infty e^{-\frac12 t^2} dt
\end{equation}

\begin{figure}[bt!]
{\samepage\columnwidth20.5pc

\hskip 0mm
\centering \epsfxsize=7.0cm\epsfbox{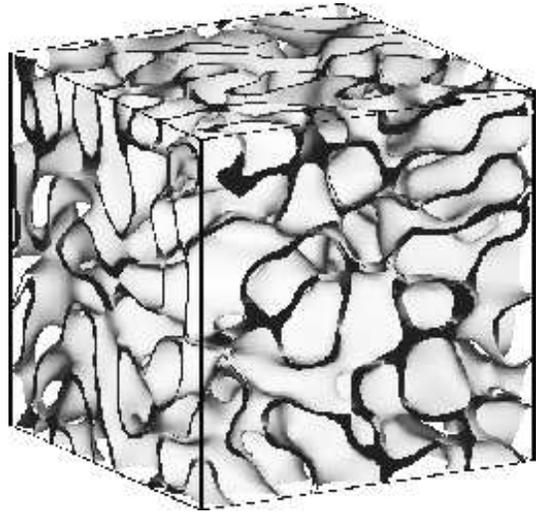}

\vspace{2mm}
\caption{\sl The interface of the media III(s=.5) at a volume fraction of
$p=0.2$. The dark region is given by $-0.253 < y({\bf r}) < 0.253$.
Note the highly connected structures.
\label{IIIs_5p_2}} } \end{figure}
\begin{figure}[bt!]
{\samepage\columnwidth20.5pc

\hskip 0mm
\centering \epsfxsize=7.0cm\epsfbox{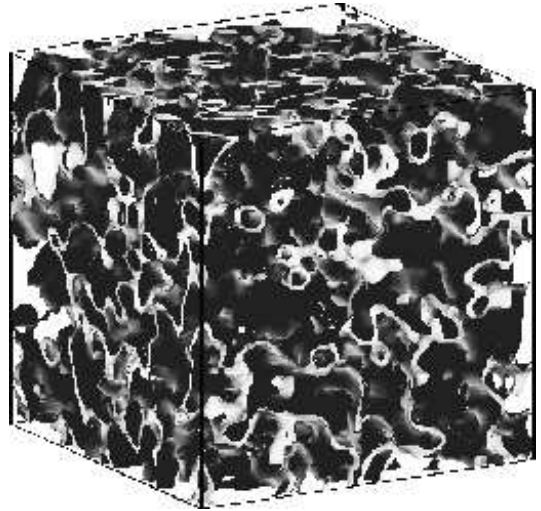}

\vspace{4mm}
\caption{\sl The interface of the media I(s=.5) with a pore volume fraction of
$p=0.2$. The light region corresponds to $-0.253 <y({\bf r}) <
0.253$.
\label{Is_5p_2i}}}
\end{figure}

\vspace{2mm}

\noindent
where $b\in[0,1]$.
Since $\beta$ is fixed, $\alpha$ is calculated using Eqn.~(\ref{ab1pnt}).
In terms of nomenclature used to describe the spectra previously these
models are denoted, for example, as III(b=.3) or I(b=0) (ie.\ $\beta=\infty$).
The latter case corresponds to the 1-level cut field.

Depending on the spectra employed and the choice of $\alpha$ and $\beta$
the 2-cut GRF scheme can model a wide range of morphologies observed in
physical composites.
The morphology of 1-cut fields is characterized by a random array of
irregular inclusions interconnected by narrower necks~\cite{Roberts95a}
similar to a `node/bond' geometry (see

\begin{figure}[bt!]
{\samepage\columnwidth20.5pc

\hskip 0mm
\centering \epsfxsize=7.0cm\epsfbox{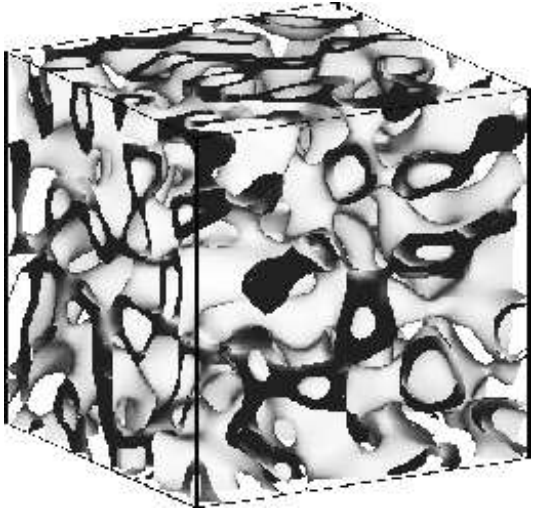}

\vspace{2mm}
\caption{\sl The interface of the media III(s=.2) at a volume fraction of
$p=0.2$. The dark region is given by $-1.28  < y({\bf r}) < -0.253$.
\label{IIIs_2p_2}}} \end{figure}

\vspace{4mm}

\noindent
Fig.~\ref{IIIb_0p_2}(c)).
This type of pore/inclusion shape has been observed in
a range of materials including alloys~\cite{Li92} and
sedimentary rocks~\cite{Bernabe91}. 
Taking $\beta\!\!=\!\!-\alpha$ in the 2-cut model $(s=0.5)$ leads to
`sheet-like' structures (see Figs.~\ref{IIIs_5p_2} \&~\ref{Is_5p_2i}) with differing degrees of roughness.
The smooth-sheet like structures of model III(s=.5) (Fig.~\ref{IIIs_5p_2})
are similar to the pores observed in dolomitic limestone~\cite{Wardlaw76},
and the connected matrix in solid foams~\cite{Roberts95b}
(see Fig.~\ref{polysty}) and polymer blends~\cite{Knackstedt95a}.
The rough `sheet-like' morphology evident in model I(s=.5) ($K=8$)
(Fig.~\ref{Is_5p_2i}) is similar to the rough porous structures observed in
pore-cast studies of sandstones~\cite{Adler92,Bernabe91}. 
Note that certain classes of sandstone have been
shown to have a fractal pore surface with $D_s\approx 2.5$~\cite{Wong86}.
This can be modelled by taking $K\to\infty$ in spectrum I
(Appx.~\ref{surfacefd}).
Qualitatively different microstructures
can be obtained in the 2-cut scheme if $\beta\neq\alpha$. For example,
the morphology of Model III(s=.2) (Fig.~\ref{IIIs_2p_2}) has both
a node/bond and sheet-like quality.

%
\section{Overlapping hollow spheres}
\label{iosa}
A second low porosity model can be defined by generalizing
the `Identical Overlapping Sphere' (IOS) model to the case
of overlapping annuli (IOSA).  
For this model the probability that $n$ points
${\mathbf r}_i$ chosen at random will fall in the {\it void} phase (ie.\
outside the hollow spheres) is just
\begin{equation} q_n=\exp[-\rho V\!a^{(n)}_U({\mathbf r}_1,\dots,{\mathbf r}_n)]. \end{equation}
Here $V\!a^{(n)}_U$ is the union volume of $n$ spherical annuli with centers
at ${\mathbf r}_i$, and $\rho$ is the number density of the annuli.

To see this consider a large region of the composite material of
volume $V$ which contains $N=\rho V$ randomly
positioned (i.e.\ uncorrelated) spherical annuli.
Now consider $q_n$ defined above.
If, and only if, the center of an annulus is
located within the volume $V\!a^{(n)}_U$,
then one (or more) of the $n$ points will lie in the solid phase.
Since each annulus is uniformly distributed
the probability that its center will not fall in the volume
$V\!a^{(n)}_U$ is $(1-V\!a^{(n)}_U / V)$.
Now there are $N$ such uncorrelated spheres so
\begin{equation}
q_n = \lim_{N\to\infty} \left({1-\frac{\rho V\!a^{(n)}_U}{N}}\right)^N =
\exp\left({-\rho V\!a^{(n)}_U}\right)
\end{equation}
where $V$, and hence $N$, has been taken to be infinitely
large. This argument (for the spherical case) is due to
\mbox{W.\ F.\ Brown}~\cite{Weissberg63,TorqStel83}.
By definition $q_n$ is just the $n$-point
void-void correlation function.
To distinguish the correlation functions associated with the
void and solid we refer to above model as the inverse IOSA model
(as the correlation function corresponds to the phase outside the annuli).
The correlation
functions for the IOSA model ($p_n$) are then just linear combinations
of $q_n$, $q_{n-1}$ etc. For example, $p_1=1-q_1$ and
$p_2(r_{12})=1-2q+q_2(r_{12}).$

Suppose the inner and outer radii of the annuli are $\mu$ and $\nu$, then
the union volume of a single annulus is 
$V\!a=V\!a^{(1)}_U= 4\pi(\nu^3-\mu^3)/3$. 
The number density of the annuli is related to the volume fraction of
void ($q$) by the formula $\rho=-{\log q}/{V\!a}$.
The higher order union volumes are derived in terms of the intersection
volumes of spheres of different radii.
For the union volume of two annuli a distance $d$ apart we have
\begin{equation} V\!a^{(2)}_U(d)=2 V\!a-V\!a^{(2)}_{I}(d) \end{equation}
where $V\!a^{(2)}_{I}(d)$ is the intersection volume of 2 annuli.
This function is given by
\begin{equation} \label{iosa2} V\!a^{(2)}_I(d)=
V^{(2)}_{I\nu\nu}(d)+V^{(2)}_{I\mu\mu}(d)-2V^{(2)}_{I\mu\nu}(d) \end{equation}
with $V^{(2)}_{Ixy}(d)$ the intersection volume of two spheres of
radii $x$ and $y$ (see Appendix \ref{ivols}).
The union volume of the three annuli distances $a$, $b$ \& $c$ apart is
\begin{eqnarray} \nonumber
V\!a^{(3)}_U(a,b,c)&=&3 V\!a- V\!a^{(2)}_I(a)- V\!a^{(2)}_I(b)- V\!a^{(2)}_I(c)\\
&& + V\!a^{(3)}_I(a,b,c),
\end{eqnarray}
where the intersection volume of three annuli ($V\!a^{(3)}_I$) is
\begin{eqnarray} \label{iosa3} \nonumber V\!a^{(3)}_I&=&
V^{(3)}_{I\nu\nu\nu}-V^{(3)}_{I\mu\mu\mu} 
- V^{(3)}_{I\mu\nu\nu}-V^{(3)}_{I\nu\mu\nu}-V^{(3)}_{I\nu\nu\mu} \\ && 
+V^{(3)}_{I\mu\mu\nu}+V^{(3)}_{I\mu\nu\mu}+ V^{(3)}_{I\nu\mu\mu}.
\end{eqnarray}
Here the function $V^{(3)}_{Ixyz}(a,b,c)$ is the intersection volume of three
spheres of radii $x$, $y$ and $z$ with $a$ the distance between the spheres
of radii $y\;\&\;z$, $b$ the distance between the spheres
of radii $x\;\&\;z$ and $c$ the distance between the spheres $x\;\&\;y$
(see Appendix \ref{ivols}).

\begin{figure}[bt!]
{\samepage\columnwidth20.5pc

\hskip 0mm
\centering \epsfxsize=7.0cm\epsfbox{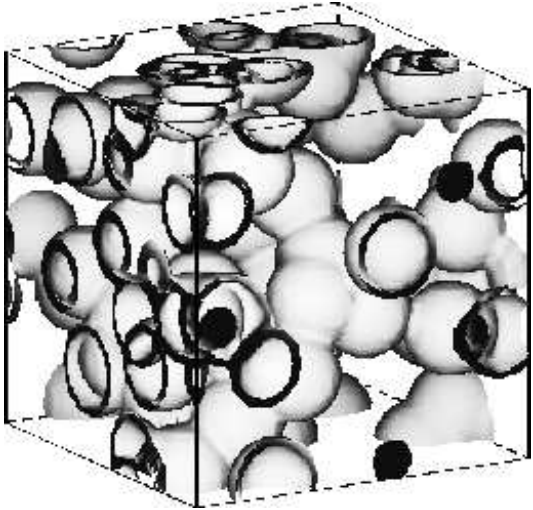}

\vspace{2mm}
\caption{\sl The interface of the IOSA model ($r_0=0.8$, $r_1=1.0$) 
at a volume fraction of $p=0.2$.  \label{IOSAr_8p_2}}} \end{figure}


As in the 2-level cut GRF model there are two obvious ways of
choosing the internal ($r_0=\mu$) and external ($r_1=\nu$) radii 
for a given volume fraction.
In the first the internal radii of the
spheres is held fixed and the number density of spheres is
increased to achieve a given volume fraction. This model morphology
corresponds to manufactured materials comprised of sintered
similarly sized hollow spheres~\cite{Green85}.
A plot of the interface for the IOSA model is given for
the case $r_0=0.8$ and $p=0.2$ in Fig.~\ref{IOSAr_8p_2}.
Using results~\cite{TorqRev91,Chiew84,Stell87,Joslin86}
developed for overlapping solid spheres (ie.\ IOS) it is possible
to incorporate a distribution of sphere sizes in the hollow sphere model.
However, polydispersity effects have been shown to be quite
small~\cite{Thovert90}.
In the second model the number density of spheres is held
fixed (so that the maximum volume fraction achievable is $p_{max}$)
and the internal radii is varied to achieve a given volume fraction.

The percolation thresholds of each phase of the IOSA model
$p_c^a$ (solid) \& $q_c^a$ (void), can be easily derived from
a knowledge of the threshold values of the
standard IOS model: $p_c^s\approx 0.3$~\cite{Pike74}
and $q_c^s\approx0.03$~\cite{Kertesz81}.
For the variable density model ($r_0$ fixed) the percolation
thresholds are $p_c^a=1-(1-p_c^s)^{1-(r_0/r_1)^3}$ and 
$q_c^a=(q_c^s)^{1-(r_0/r_1)^3}$ (so $p_c^a\to0$ \& $q_c^a\to1$ 
as $r_0\to r_1$).
For the fixed density model the IOSA solid
phase is percolative if $p_{max} \geq p_c^s$ and the void phase is
percolative if $q_{min}=1-p_{max}\geq q_c^s$. 
\section{Microstructure parameters}
\label{zeta_ab}
Bounds have been calculated on the conductivity~\cite{Beran65a,Milton81b}
and the bulk~\cite{Beran65b} and shear~\cite{McCoy,Milton82b} moduli
of composite materials (reviewed in Ref.~\cite{TorqRev91}).
These can be expressed~\cite{Milton81a} in terms of the
volume fractions and properties of each of the phases and two
microstructure parameters:
\begin{eqnarray}
\label{zeta}
\zeta_1&=& 
\frac9{2pq}\int_0^\infty\!\!\frac{dr}{r} \int_0^\infty
\!\! \frac{ds}{s} \int_{-1}^1 \!\! du  P_2(u) f(r,s,t) \\
\eta_1&=&\frac {5\zeta_1}{21}+
\frac{150}{7pq}\int_0^\infty\!\!\frac{dr}{r}
\int_0^\infty\!\! \frac{ds}{s}
\int_{-1}^1 \!\! du  P_4(u) f(r,s,t) 
\label{eta}
\end{eqnarray}    
where $f(r,s,t)=p_3(r,s,t)-p_2(r)p_2(s)/p$, $t^2=r^2+s^2-2rs u$
and $P_n(u)$ denotes the Legendre polynomial of order $n$.
As we argued in Ref.~\cite{Roberts95a} it only appears necessary
to know broad microstructural information about a general
composite to successfully apply the bounds.
This conclusion arose from the observation that
the bounds are relatively insensitive to small variations in the
microstructure parameters. Furthermore we found that the parameters
$\zeta_1$ and 
$\eta_1$ are insensitive to fine microstructural details within a class
of composites (e.g.\ the overlapping sphere class, or the 1-level
cut GRF class). An
example of this insensitivity is also seen when polydispersity effects of
particulate models are considered~\cite{Thovert90}.
In light of these facts the parameters calculated from
models may well have application to physical composites
for which precise microstructural information is unavailable.

\begin{figure}[bt!]
{\samepage\columnwidth20.5pc

\hskip 0mm
\centering \epsfxsize=8.5cm\epsfbox{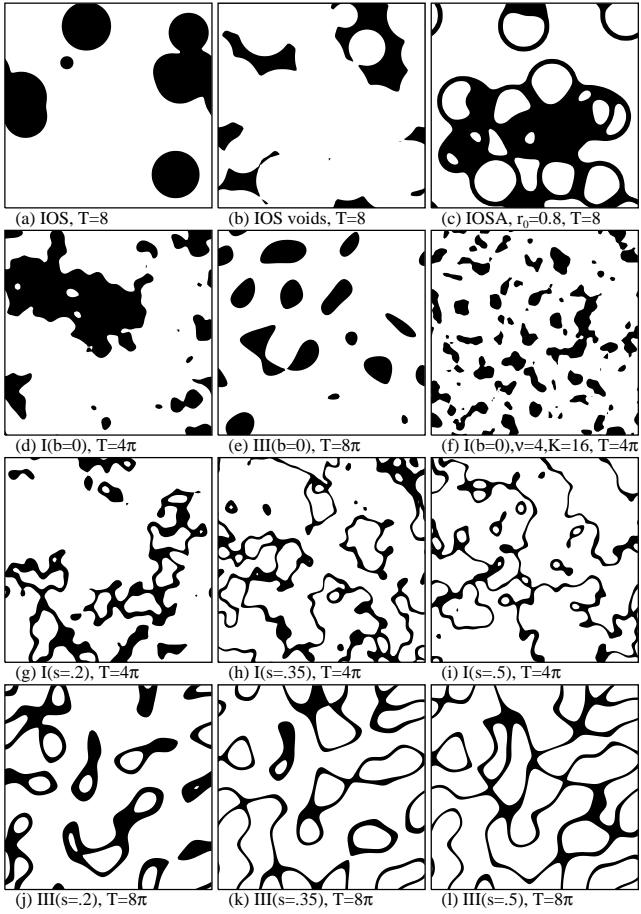}

\vspace{2mm}
\caption{\sl A variety of microstructures (volume fraction 20\%) 
for which the third order statistics are
known exactly:
(a)-(c) IOS and IOSA models ($r_1=1$);
(d)-(f) 1-level cut GRF's;
(g)-(l) 2-level cut GRF's.
Except where noted Model I has $K=8$ \& $\nu=0$ and Model III $\mu=1.5$.
\label{modelsab}}}
\end{figure}

In Fig.~\ref{modelsab} we provide a graphical summary of 
the wide range of isotropic composites for which $p_3$
(and hence the microstructure parameters) can been exactly calculated. 
It is clear that the 2-cut GRF and overlapping hollow sphere
model considerably expand the classes of materials to which the bounds can
be applied.

We now report calculations of the microstructure
parameters for a variety of 2-level cut GRF and IOSA models.
Our method of calculating $\zeta_1$ (and $\eta_1$) has been
discussed previously~\cite{Roberts95a}. In addition we employ
an adaptive integration algorithm~\cite{Roberts95d} to
compensate for the fact that the sub-integrand
$\int_{-1}^1P_n(u)f(r,s,t)du$ varies rapidly in the region $r\approx0$
and involves a considerable number of function evaluations. 
The error in the results is less than 1\%.
To model as wide a range of materials as possible three
qualitatively different spectra are used in the level-cut GRF scheme;
models I ($K=\infty$), I ($K=8$) and III ($\mu=1.5$).
These spectra lead to surface-fractal, rough and smooth interfaces
respectively.

\begin{figure}[bt!]
{\samepage\columnwidth20.5pc

\vspace{-4mm}

\hskip 0mm
\centering \epsfxsize=8.5cm\epsfbox{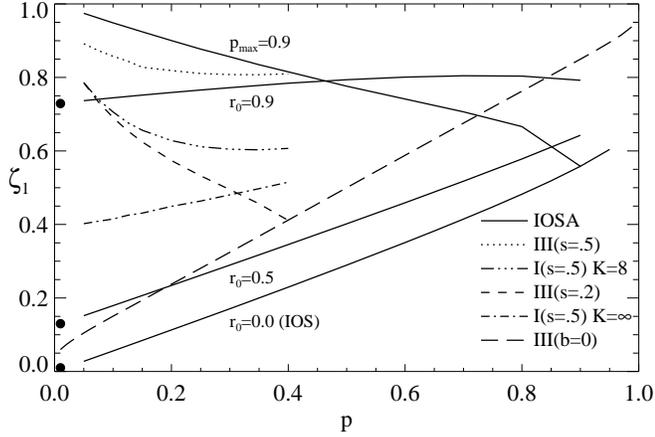}

\vspace{4mm}
\caption{\sl The microstructure parameter $\zeta_1$ for selected models.
The IOS model and model III(b=0) are included to show the behaviour of $\zeta_1$
for different classes of composites (see Fig.\ \ref{modelsab}).
The solid symbols represent analytic calculations of $\zeta_1|_{p=0}$
for the IOSA model.
\label{z_ab_sum}}}
\end{figure}

\begin{figure}[bt!]
{\samepage\columnwidth20.5pc

\hskip 0mm
\centering \epsfxsize=8.5cm\epsfbox{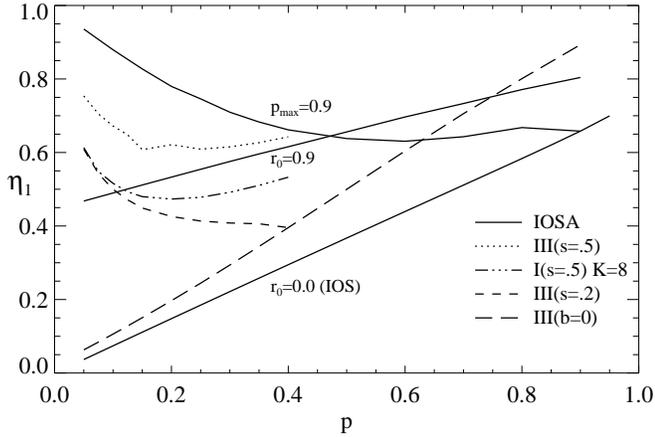}

\vspace{4mm}
\caption{\sl The elasticity microstructure parameter $\eta_1$ for selected
models.  The IOS model and model III(b=0) are included to show
the behaviour of $\eta_1$ for different classes of composites
(see Fig.\ \ref{modelsab}).
\label{e_ab_sum}}}
\end{figure}

As we are primarily interested in low volume fraction porous or
solid media the microstructure parameters we report are 
in the range $0.0 < p \leq 0.4$. The results for $\zeta_1$ and $\eta_1$ 
are given in Tables~\ref{tab_z_ab_s}~\&~\ref{tab_e_ab_s} and selected
results are plotted in Figs.~\ref{z_ab_sum}~\&~\ref{e_ab_sum}.
The results for the two variants of the IOSA model are given in
Table~\ref{tab_iosa} and plotted along side the results
for the 2-cut GRF models in Figs.~\ref{z_ab_sum}~\&~\ref{e_ab_sum}.
Due to the simple geometry of the IOSA model it is
possible to calculate $\sigma_e$ to order $p$
(see Appendix \ref{confzeta}).
This result can then be used to show $\zeta_1|_{p=0}=(r_0/r_1)^3$
(represented by symbols in Fig.~\ref{z_ab_sum}) in agreement with our numerical
calculations of $\zeta_1$.

\begin{figure}[bt!]
{\samepage\columnwidth20.5pc

\hskip 0mm
\centering \epsfxsize=7.6cm\epsfbox{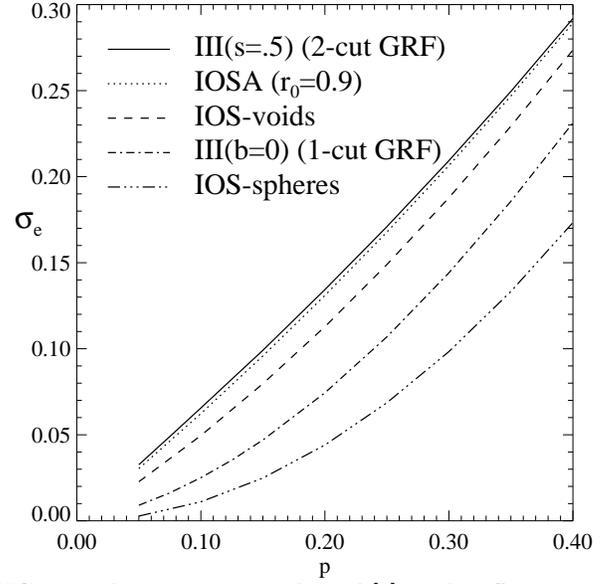}

\vspace{3mm}
\caption{\sl The 3-point upper bound~\protect\cite{Beran65a}
on the effective conductivity (contrast 1:0) 
of five different classes of
microstructure. The data for the IOS and 1-cut GRF models are
from Refs.~\protect\cite{TorqStel83,Roberts95a}.  
\label{sigmae}}}
\end{figure}

To compare the properties of different media we
plot (Fig.~\ref{sigmae}) the upper bound on the
conductivity for one member of each class of
composite: 2-cut GRFs, hollow spheres,
IOS-voids~\cite{TorqStel83} (or swiss-cheese),
1-cut GRFs~\cite{Roberts95a} and IOS~\cite{TorqStel83} (or solid spheres). 

\noindent
\begin{minipage}[b!]{8.5cm}
\begin{table}
\caption{The microstructure parameter $\zeta_1$ for a range of
materials generated from the symmetric GRF model.}
\label{tab_z_ab_s}
\begin{tabular}{cccccccccc}
\multicolumn{1}{c}{Mod.\ } &
\multicolumn{3}{c}{I, $K=\infty$} &
\multicolumn{3}{c}{I, $K=8$} &
\multicolumn{3}{c}{III} \\
\hline
s  &  .20 &   .35 &   .50 &   .20 &   .35 &   .50 &.20&.35&.50 \\
\hline
\multicolumn{1}{c}{p} &
\multicolumn{9}{c}{$\zeta_1$} \\
\hline
.050& .401 & .401 & .402 & .706 & .773 & .786 & .785 & .872 & .892 \\
.075& .402 & .408 & .409 & .641 & .719 & .739 & .733 & .845 & .873 \\
.100& .405 & .413 & .415 & .597 & .684 & .706 & .691 & .824 & .858 \\
.125& .410 & .422 & .425 & .563 & .655 & .677 & .655 & .807 & .845 \\
.150& .417 & .428 & .431 & .536 & .633 & .656 & .625 & .791 & .828 \\
.200& .425 & .443 & .449 & .500 & .601 & .628 & .574 & .769 & .819 \\
.250& .435 & .459 & .464 & .478 & .583 & .611 & .532 & .753 & .811 \\
.300& .443 & .475 & .481 & .464 & .575 & .605 & .495 & .741 & .808 \\
.350& .451 & .491 & .497 & .455 & .572 & .603 & .456 & .734 & .807 \\
.400& .456 & .506 & .515 & .444 & .574 & .607 & .411 & .728 & .810 \\
\end{tabular} \end{table}
\end{minipage}

\noindent
\begin{minipage}[b!]{8.5cm}
\begin{table}
\caption{The elasticity microstructure parameter $\eta_1$ for a range of
materials generated using the symmetric GRF model.}
\label{tab_e_ab_s}
\begin{tabular}{ccccccc}
\multicolumn{1}{c}{Mod.\ } &
\multicolumn{2}{c}{I, $K=\infty$} &
\multicolumn{2}{c}{I, $K=8$} &
\multicolumn{2}{c}{III} \\
\hline
 s   &.20 & .50 & .20 & .50 & .20 & .50 \\
\hline
\multicolumn{1}{c}{p} &
\multicolumn{6}{c}{$\eta_1$} \\
\hline
.050& .355 & .351 & .523 & .613 & .609 & .754 \\
.075& .358 & .362 & .463 & .548 & .543 & .705 \\
.100& .362 & .369 & .430 & .516 & .500 & .672 \\
.125& .370 & .377 & .416 & .493 & .471 & .648 \\
.150& .373 & .388 & .407 & .480 & .449 & .608 \\
.200& .394 & .402 & .404 & .473 & .426 & .621 \\
.250& .410 & .430 & .410 & .478 & .414 & .609 \\
.300& .426 & .451 & .420 & .492 & .408 & .615 \\
.350& .438 & .474 & .430 & .510 & .406 & .627 \\
.400& .442 & .495 & .431 & .533 & .396 & .643 \\
\end{tabular} \end{table}

\vspace{-4mm}

\noindent
\begin{table}
\caption{The microstructure parameters for different versions of the
IOSA model. For the case $r_0=0$ the results are just those of the
standard IOS model (see Ref.~\protect\cite{TorqRev91}).}
\label{tab_iosa}
\label{tabboundsI}
\begin{tabular}{ccccccccc}
\multicolumn{1}{c}{Mod.\ }   &
\multicolumn{2}{c}{$r_0=0.5$}   &
\multicolumn{2}{c}{$r_0=0.9$}   &
\multicolumn{2}{c}{$p_{max}=0.9$}   &
\multicolumn{2}{c}{$p_{max}=0.7$}   \\
\hline
\multicolumn{1}{c}{p} &
\multicolumn{1}{c}{$\zeta_1$} &\multicolumn{1}{c}{$\eta_1$} &
\multicolumn{1}{c}{$\zeta_1$} &\multicolumn{1}{c}{$\eta_1$} &
\multicolumn{1}{c}{$\zeta_1$} &\multicolumn{1}{c}{$\eta_1$} &
\multicolumn{1}{c}{$\zeta_1$} &\multicolumn{1}{c}{$\eta_1$} \\
\hline
.05 & .152 & .119 & .737 & .468 & .974 & .936 & .955 & .888 \\
.10 & .179 & .153 & .744 & .490 & .948 & .880 & .911 & .788 \\
.15 & .207 & .187 & .752 & .512 & .924 & .827 & .870 & .707 \\
.20 & .234 & .221 & .759 & .533 & .900 & .780 & .829 & .640 \\
.25 & .262 & .254 & .766 & .554 & .877 & .746 & .791 & .590 \\
.30 & .289 & .288 & .772 & .576 & .856 & .710 & .754 & .551 \\
.35 & .317 & .322 & .778 & .596 & .835 & .683 & .718 & .524 \\
.40 & .345 & .356 & .784 & .616 & .815 & .662 & .683 & .508 \\
.50 & .402 & .424 & .794 & .656 & .776 & .638 & .614 & .502 \\
.60 & .459 & .492 & .801 & .697 & .741 & .631 & .539 & .520 \\
.70 & .517 & .560 & .805 & .733 & .706 & .643 & .414 & .511 \\
.80 & .578 & .630 & .804 & .771 & .666 & .668 &       &       \\
.90 & .643 & .705 & .792 & .804 & .558 & .658 &       &       \\
\end{tabular}
\end{table}
\end{minipage}

\begin{figure}[bt!]
{\samepage\columnwidth20.5pc

\hskip 0mm
\centering \epsfxsize=7.6cm\epsfbox{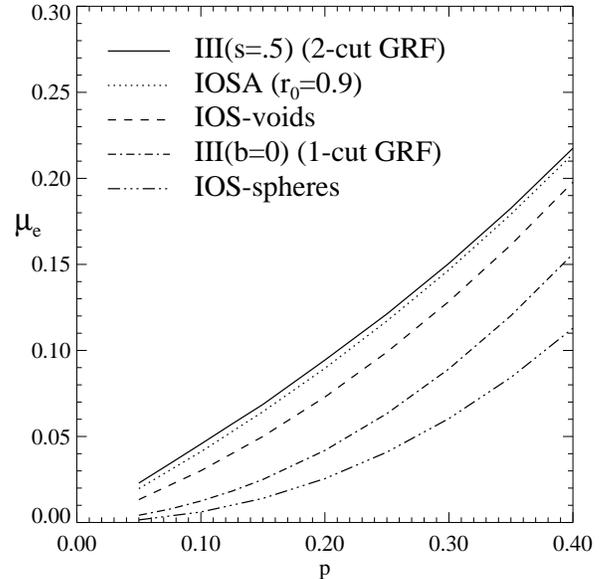}

\vspace{3mm}
\caption{\sl The 3-point upper bound~\protect\cite{Milton82b}
on the effective shear modulus (contrast 1:0) of five different classes of
microstructure. The data for the IOS and 1-cut GRF models are
from Refs.~\protect\cite{TorqStel83,Roberts95a}.
\label{mue}}}
\end{figure}

We have also evaluated bounds on the shear, bulk and Young's moduli of
the models. In Ref.~\cite{Roberts95b} we showed that the upper bound on Young's
modulus was in good agreement with experimental measurements for foamed
solids. Model III(s=.5) provides a good model of polystyrene foam
(compare Figs.~\ref{polysty}~\&~\ref{IIIs_5p_2}), and the IOSA
model accurately mimics the microstructure and properties of sintered hollow
glass spheres.
In Fig.~\ref{mue} the upper bound on the shear modulus is shown for
each class of composite considered above: the microstructure clearly
has a strong influence on elastic properties. The bulk and Young's moduli
show similar behaviour. 

\vspace{-2mm}
  
\section{Simulations of $\sigma_e$}
\label{simuls_ab}

\vspace{-2mm}

In addition to bounding the properties of composite media
and providing qualitative information on these properties, it has been
observed that the bounds also have reasonable predictive
power~\cite{TorqRev91}. To test the
predictive utility of the bounds
and provide a direct

\vspace{-11mm}

\vspace{-1mm}

\noindent
comparison between microstructure and properties
we use a finite-difference method to explicitly calculate the
conductivity of several 2-cut GRF's.

The effective conductivity $\sigma_e$ of a composite is defined as
the ratio of the current density to the applied potential.
We take $T$ as the scale of the sample and
$M^3$ as the number of nodes (so the spatial resolution
scale is $\Delta x = T/M$).
The generation of random fields and the method for determining
$\sigma_e$ were described in Ref.~\cite{Roberts95a} for the case of
1-cut fields. 
A number of additional difficulties are encountered in the simulations
of $\sigma_e$ for the 2-level cut GRF's. The major problems are;
(i) {\it discretisation} effects which occur when the discretisation
length scale $\Delta x$ is insufficient to resolve the thin sheet-like
structures which arise (e.g.\ Fig.~\ref{IIIs_5p_2}) and (ii) 
{\it finite-scale} effects which arise if $T$ is not large enough to
represent an `infinite'
medium. In practice $T$ should be several times the

\vspace{-8mm}

\vspace{-4mm}

\noindent
correlation
length of the microstructure (approx.\ unity).
Discretisation effects can be reduced by increasing $M$ or
decreasing $T$ (to increase the width of the sheets relative to
$\Delta x$). However our computational requirements dictate $M\leq128$ and
decreasing $T$ leads to noisy results.
Thus $T$ must be chosen to minimize each of these
competing errors. By performing several numerical
tests~\cite{Roberts95d} a reasonable value of $T$
was determined to ensure that simulations of $\sigma_e$ are accurate.
As the sheets become

\end{multicols}
\begin{multicols}{2}

\begin{figure}[bt!]
{\samepage\columnwidth20.5pc

\hskip 0mm
\centering \epsfxsize=7.5cm\epsfbox{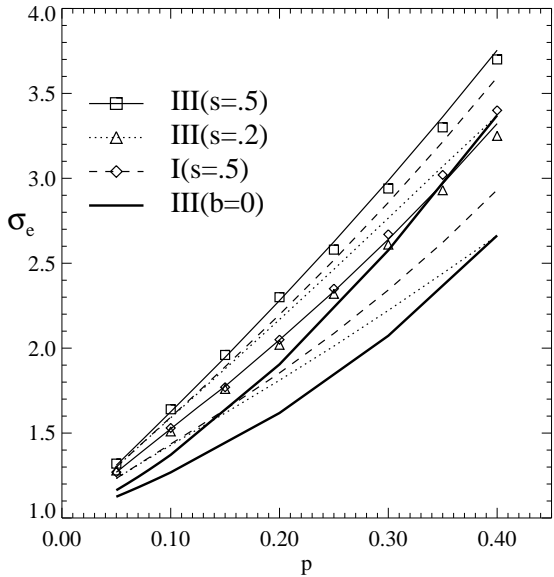}

\vspace{4mm}
\caption{\sl
The simulation data and bounds for three different 2-level
cut GRF models at contrast $\sigma_{1,2}=10,1$. The thick solid line
corresponds to the bounds for the 1-level cut model III(b=0).
The bounds are clearly seen to differentiate the different classes
of media at low $p$.
\label{ab10_1}}}
\end{figure}

\vspace{-5mm}

\noindent
\begin{minipage}[b!]{8.5cm}
\begin{table}
\caption{The effective conductivity of several 2-level cut GRF's for the
case $\sigma_{1,2}=10,1$.}
\label{tab2_10_1}
\begin{tabular}{cccccccccc}
\multicolumn{1}{c}{}   &
\multicolumn{3}{c}{III(s=.2)}   &
\multicolumn{3}{c}{III(s=.5)}   &
\multicolumn{3}{c}{I(s=.5)}   \\
\hline
\multicolumn{1}{c}{p} &
\multicolumn{1}{c}{$T/\pi$} &\multicolumn{1}{c}{$\sigma_e$} &
\multicolumn{1}{c}{Err.} &
\multicolumn{1}{c}{$T/\pi$} &\multicolumn{1}{c}{$\sigma_e$} &
\multicolumn{1}{c}{Err.} &
\multicolumn{1}{c}{$T/\pi$} &\multicolumn{1}{c}{$\sigma_e$} &
\multicolumn{1}{c}{Err.} \\
\hline
0.05 &4& 1.28 & .01 &2& 1.32 & .02 &2& 1.27  & .00 \\
0.10 &4& 1.51 & .02 &2& 1.64 & .03 &2& 1.53  & .02 \\
0.15 &4& 1.76 & .04 &2& 1.96 & .05 &2& 1.77  & .02 \\
0.20 &4& 2.02 & .06 &4& 2.30 & .07 &2& 2.05  & .01 \\
0.25 &8& 2.32 & .03 &4& 2.58 & .10 &4& 2.35  & .01 \\
0.30 &8& 2.61 & .04 &4& 2.94 & .12 &4& 2.67  & .01 \\
0.35 &8& 2.93 & .03 &4& 3.30 & .13 &4& 3.02  & .01 \\
0.40 &8& 3.25 & .05 &4& 3.70 & .13 &4& 3.40  & .02 \\
\end{tabular} \end{table}
\end{minipage}

\begin{figure}[bt!]
{\samepage\columnwidth20.5pc

\hskip 0mm
\centering \epsfxsize=7.5cm\epsfbox{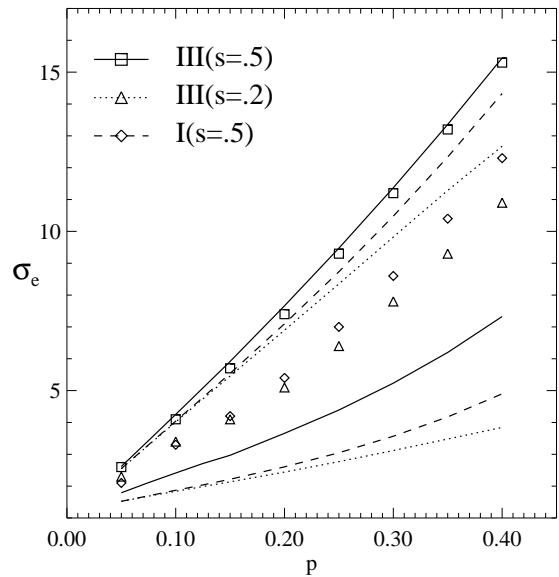}

\vspace{4mm}
\caption{\sl 
The simulation data and bounds for three different 2-level
cut GRF models with conductivity $\sigma_{1,2}=50,1$. 
\label{ab50_1}}}
\end{figure}

\vspace{-5mm}

\noindent
\begin{minipage}[b!]{8.5cm}
\begin{table}
\caption{The effective conductivity of several 2-level cut GRF's for the
case $\sigma_{1,2}=50,1$.}
\label{tab2_50_1}
\begin{tabular}{cccccccccc}
\multicolumn{1}{c}{}   &
\multicolumn{3}{c}{III(s=.2)}   &
\multicolumn{3}{c}{III(s=.5)}   &
\multicolumn{3}{c}{I(s=.5)}   \\
\hline
\multicolumn{1}{c}{p} &
\multicolumn{1}{c}{$T/\pi$} &\multicolumn{1}{c}{$\sigma_e$} &
\multicolumn{1}{c}{Err.} &
\multicolumn{1}{c}{$T/\pi$} &\multicolumn{1}{c}{$\sigma_e$} &
\multicolumn{1}{c}{Err.} &
\multicolumn{1}{c}{$T/\pi$} &\multicolumn{1}{c}{$\sigma_e$} &
\multicolumn{1}{c}{Err.} \\
\hline
0.05 &4& 2.3 & 0.2 &2& 2.6 & 0.2 &1& 2.1 &0.1 \\
0.10 &4& 3.4 & 0.4 &2& 4.1 & 0.4 &1& 3.3 &0.1 \\
0.15 &8& 4.1 & 0.1 &2& 5.7 & 0.5 &1& 4.2 &0.1 \\
0.20 &8& 5.1 & 0.2 &2& 7.4 & 0.7 &1& 5.4 &0.1 \\
0.25 &8& 6.4 & 0.2 &4& 9.3 & 0.2 &2& 7.0 &0.1 \\
0.30 &8& 7.8 & 0.2 &4&11.2 & 0.3 &2& 8.6 &0.1 \\
0.35 &8& 9.3 & 0.3 &4&13.2 & 0.3 &2&10.4 &0.2 \\
0.40 &8& 10.9& 0.4 &4&15.3 & 0.3 &2&12.3 &0.2 \\
\end{tabular} \end{table}
\end{minipage}

\end{multicols}

\vspace{-10mm}
$\;$

\begin{multicols}{2}

\noindent
thinner (ie.\ $p$ decreases)
it was found that smaller values of $T$ are necessary
to eliminate finite-scale effects. This can be explained in terms
of the faster decay of the correlations between
the components of phase 1.

We choose to study the effective conductivity of models
III(s=.5), I(s=.5) ($K=8$) and III(s=.2). 
The former models provide examples
of smooth and rough `sheet-like' pores. The latter model (III(s=.2)) has
a morphology comprised of inclusions with both a sheet- and node/bond-like
quality. 
The conductivity contrasts employed
occur in physical composites and have been studied previously,
allowing comparisons to be made.
In each of the cases we report results averaged over five samples
for a range of volume fractions $0.05 \leq p \leq 0.4$.
In all cases the simulational data lie between the bounds.

First we consider the conductivity contrast $\sigma_{1,2}=10,1$. 
The results are tabulated in Table \ref{tab2_10_1}, and
plotted in Fig.~\ref{ab10_1} along with the bounds for each model, Here,
and in subsequent calculations, the lower bound of Beran~\cite{Beran65a}
and the upper bound of Milton~\cite{Milton81b} (see Ref.~\cite{Roberts95a})
are employed.
The data for model III(s=.5) practically lies along the
relevant upper bound. In contrast the effective conductivity of
models III(s=.2) and I(s=.5) fall between the bounds,
however the upper bound still provides a reasonable estimate of
$\sigma_e$ in each case.
For purposes of comparison the bounds for
the 1-level cut GRF model III(b=0) are included in Fig.~\ref{ab10_1}.
At low $p$ the bounds clearly differentiate between the
different classes of media.  
It is clear that the model III(s=.5) is a significantly more efficient
conductor than models III(s=.2) or I(s=.5) and those defined using
a 1-level cut GRF in \cite{Roberts95a}. 
 
The simulation data for the contrast $\sigma_{1,2}=50,1$
is reported in Table \ref{tab2_50_1} and plotted in Fig.~\ref{ab50_1}.
Qualitatively the results are the same as those
discussed in relation to the case $\sigma_{1,2}=10,1$. Note that 
the upper bound is again a good estimate for model III(s=.5);
less so for models

\end{multicols}
\begin{multicols}{2}

\begin{figure}[bt!]
{\samepage\columnwidth20.5pc

\hskip 0mm
\centering \epsfxsize=7.5cm\epsfbox{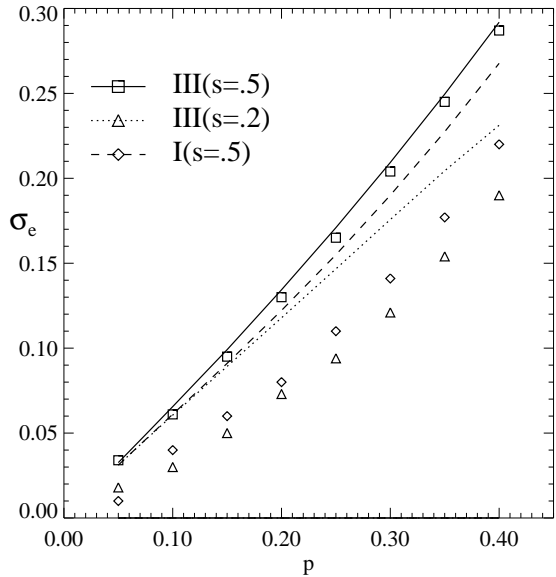}

\vspace{4mm}
\caption{\sl Simulation data and bounds for conductivity contrast
$\sigma_{1,2}=1,0$. The models and contrast are relevant to
solid-foams~\protect\cite{Roberts95b} and porous
rocks~\protect\cite{pRoberts95c}. \label{ab1_0}}}
\end{figure}

\vspace{-4mm}

\noindent
\begin{minipage}[b!]{8.5cm}
\begin{table}
\caption{The effective conductivity of several 2-level cut GRF's for the
case $\sigma_{1,2}=1,0$. \label{tab2_1_0}}
\begin{tabular}{cccccccccc}
\multicolumn{1}{c}{}   &
\multicolumn{3}{c}{III(s=.2)}   &
\multicolumn{3}{c}{III(s=.5)}   &
\multicolumn{3}{c}{I(s=.5)}     \\
\hline
\multicolumn{1}{c}{p} &
\multicolumn{1}{c}{$T\!/\pi$} &\multicolumn{1}{c}{$\sigma_e$} &
\multicolumn{1}{c}{Err.} &
\multicolumn{1}{c}{$T\!/\pi$} &\multicolumn{1}{c}{$\sigma_e$} &
\multicolumn{1}{c}{Err.} &
\multicolumn{1}{c}{$T\!/\pi$} &\multicolumn{1}{c}{$\sigma_e$} &
\multicolumn{1}{c}{Err.} \\
\hline
.05 &4& .018 & .002 &2& .034 &.003&2& .011&    .003 \\
.10 &4& .030 & .004 &2& .061 &.003&2& .044&    .005 \\
.15 &4& .050 & .005 &2& .095 &.003&2& .058&    .007 \\
.20 &4& .073 & .007 &4& .130 &.004&2& .078&    .009 \\
.25 &8& .094 & .004 &4& .165 &.005&4& .111&    .004 \\
.30 &8& .121 & .003 &4& .204 &.005&4& .141&    .004 \\
.35 &8& .154 & .003 &4& .245 &.006&4& .177&    .007 \\
.40 &8& .190 & .002 &4& .287 &.007&4& .219&    .006 
\end{tabular}
\end{table}
\end{minipage}

\begin{figure}[bt!]
{\samepage\columnwidth20.5pc

\hskip 0mm
\centering \epsfxsize=7.5cm\epsfbox{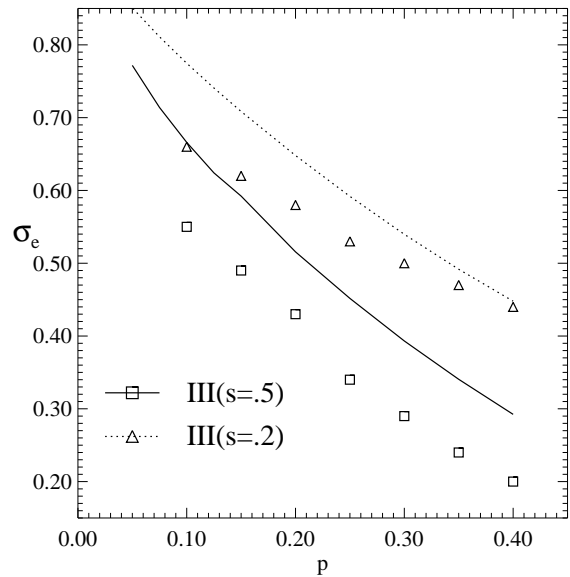}

\vspace{4mm}
\caption{\sl Simulations of $\sigma_e$ and the bounds for $\sigma_{1,2}=0,1$.
This contrast is relevant to modeling transport in membranes.
\label{ab0_1}}}
\end{figure}

\vspace{-4mm}

\noindent
\begin{minipage}[b!]{8.5cm}
\begin{table}
\caption{The effective conductivity of several 2-level cut GRF's for the
case $\sigma_{1,2}=0,1$.}
\label{tab2_0_1}
\begin{tabular}{ccccccc}
\multicolumn{1}{c}{}   &
\multicolumn{3}{c}{III(s=.2)}   &
\multicolumn{3}{c}{III(s=.5)}   \\
\hline
\multicolumn{1}{c}{p} &
\multicolumn{1}{c}{$T/\pi$} &\multicolumn{1}{c}{$\sigma_e$} &
\multicolumn{1}{c}{Err.} &
\multicolumn{1}{c}{$T/\pi$} &\multicolumn{1}{c}{$\sigma_e$} &
\multicolumn{1}{c}{Err.} \\
\hline
0.10 &4& 0.66 & .02 &2& 0.55 & .05 \\
0.15 &4& 0.62 & .02 &2& 0.49 & .06 \\
0.20 &4& 0.58 & .02 &2& 0.43 & .07 \\
0.25 &8& 0.53 & .01 &4& 0.34 & .05 \\
0.30 &8& 0.50 & .01 &4& 0.29 & .05 \\
0.35 &8& 0.47 & .01 &4& 0.24 & .04 \\
0.40 &8& 0.44 & .01 &4& 0.20 & .03 \\
\end{tabular}
\end{table}
\end{minipage}

\end{multicols}

\vspace{-10mm}
$\;$

\begin{multicols}{2}

\noindent
III(s=.2) and I(s=.5). 

In porous rocks and solid foams the conductivity of the medium surrounding
the conducting pathways has negligible (or zero) conductivity. To model
such systems the contrast $\sigma_{1,2}=1,0$ is used.
The data and computational parameters used in
the simulations are reported in Table \ref{tab2_1_0}.
Each material is seen to be conductive at the lowest volume
fraction considered $p=0.05$.
Discretisation effects prohibit accurate simulations of $\sigma_e$
at lower volume fractions. 
The simulation
data and the upper bounds are plotted in Fig.~\ref{ab1_0}.
Even in this large contrast situation the upper bound for model
III(s=.5) agrees with the data. 

To consider the case of diffusive transport in membranes,
we assume that the membrane has negligible diffusivity with
respect to the surrounding fluid. Therefore the contrast $\sigma_{1,2}=0,1$ is
employed.  For this system large discretisation effects
prohibit the consideration of model I(s=.5) and membrane/pore volume fractions
of less than $p=0.10$.
The data is presented in Table \ref{tab2_0_1} and Fig.~\ref{ab0_1}.
Note that the presence of a membrane
occupying 10\%-20\% of the total volume reduces the diffusivity by a
factor of two. This is due to the tortuous pathways through which the
diffusing species must migrate.
In contrast to three cases considered above the upper
bound does not provide a good estimate of $\sigma_e$ for model III(s=.5).

\vspace{-3mm}
\section{Effect of microstructure on properties}
\label{influence}

\vspace{-3mm}

The precise role of microstructure in determining the macroscopic properties
of composite media has been the subject of many studies. A number of
simple models of pore-shape have proposed to determine,
for example, the effect of pore-size distribution~\cite{Wong84},
pore roughness~\cite{Schwartz89} and pore geometry~\cite{Bernabe91} on
transport in porous rocks. Simple micro-mechanical models~\cite{Christensen90}
have also been studied to ascertain, for example, the effect of
inclusion shape~\cite{Wu66} and cell structure~\cite{Gibson88,Gibson82} 
on the mechanical properties of composites.
In this section we investigate

\end{multicols}

\begin{center}
\noindent
\begin{minipage}[b!]{16cm}
\begin{table}
\caption{Qualitative comparison of microstructure and macroscopic
properties. We consider the contrast 1:0 at a representative volume
fraction $p=0.2$. $\sigma_u$, $\kappa_u$, and $\mu_u$ are upper bounds
on the effective conductivity~\protect\cite{Beran65a}, bulk~\protect\cite{Beran65b} and
shear modulus~\protect\cite{Milton82b} respectively.}
\label{qual_table}
\begin{tabular}{llccccccc}
Model & Microstructure &
$p_c$ & $\sigma_e$ & $\zeta_1$ & $\eta_1$ & $\sigma_u$ & $\kappa_u$ & $\mu_u$ \\
\hline
III(s=.5)             &  smooth, sheet-like    &
0  & 0.130  & .819 & .621 & 0.134 & 0.115 & 0.094 \\
I(s=.5) $K=8$       &  rough, sheet-like     &
0  & 0.078  & .628 & .473 & 0.122 & 0.102 & 0.081 \\
III(s=.2)             &  smooth, node/bond/sheet-like  &
0  & 0.073  & .574 & .426 & 0.118 & 0.098 & 0.077 \\
I(s=.5) $K=\infty$  &  very rough, sheet-like &
0 & -       & .449 & .402 & 0.106 & 0.086 & 0.068 \\
I(b=0)\tablenotemark[1]$K=\infty$  &  very rough, node/bond-like &
-  & -      & .366 & .333 & 0.096 & 0.076 & 0.060 \\
I(b=0)\tablenotemark[1]$K=8$ &  rough, node/bond-like &
.07& 0.027  & .326 & .291 & 0.090 & 0.071 & 0.054 \\
III(b=0)\tablenotemark[1]             &  smooth, node/bond-like &
.13& 0.026  & .237 & .197 & 0.074 & 0.057 & 0.042 \\
\hline
IOSA ($r_0=0.9$) &  hollow spheres    &
.09& -      & .759 & .533 & 0.131 & 0.117 & 0.090 \\
IOS-voids        &  swiss-cheese    &
.03\tablenotemark[2]& 0.076\tablenotemark[3]  & .518\tablenotemark[4] & .416\tablenotemark[4] & 0.113 & 0.093 & 0.073 \\
IOS              &  spheres           &
.30\tablenotemark[5]& 0      & .113\tablenotemark[4]  & .148\tablenotemark[4] & 0.044 & 0.032 & 0.026 \\
\end{tabular}
\tablenotetext[1]{Ref.~\cite{Roberts95a}}%
\tablenotetext[2]{Ref.~\cite{Kertesz81}}%
\tablenotetext[3]{Ref.~\cite{Kim92}}%
\tablenotetext[4]{Refs.~\cite{TorqRev91,TorqStel83}}
\tablenotetext[5]{Refs.~\cite{Isichenko92,Pike74}}            
\end{table}
\end{minipage}
\end{center}

\begin{multicols}{2}

\noindent
how morphology
influences the properties of realistic model composites.

To simplify the discussion we summarize relevant data for a variety of
GRF and particulate microstructure models in Table~\ref{qual_table}.
We consider systems of 1:0 contrast at $p=0.2$. This case corresponds to
a conducting (mechanically strong) matrix in an
insulating (weak) medium (e.g.\ foamed solids~\cite{Roberts95b}).
This contrast also corresponds to low porosity
conducting pores in an insulating medium (e.g.\ porous rocks).
To gauge the effect of microstructure on material properties we assume
that the upper bound on each property provides
an estimate of its actual value.
A comparison of $\sigma_u$ and $\sigma_e$ (where available)
shows that this is generally true~\cite{prenote1}. Note, however, that
if the difference between $\sigma_u$ for each of
the models is small (e.g.\ IOS-voids and model III(s=.2) and
examples in Ref.~\cite{Roberts95a}) such an assumption cannot be
made ($\sigma_u^a > \sigma_u^b$ but $\sigma_e^a < \sigma_e^b$).

At a 1:0 contrast the effective properties only differ from zero
if the composite is macroscopically connected (i.e.\ percolative).
At $p=0.2$ this condition is satisfied for all
but one of the media (conducting spheres in an insulating medium).
Above this threshold the magnitude of the macroscopic properties
is then governed by the shape of the inclusions. 
It is clear from the table that sheet-like structures provide
higher conductivity and mechanical strength than those with
a node/bond-like character.
To elucidate the role of inclusion shape we
derive approximate expressions for the effective conductivity
of periodic media with unit cells of each type in Appendix \ref{cellfoam}.
For small volume fractions ($p\ll1$) the node/bond model has
$\sigma_e\simeq p^2\sigma_1$ and the sheet-like model has
$\sigma_e\simeq \frac23 p\sigma_1$ in qualitative agreement with the data.
Interestingly the periodic sheet model provides a surprisingly
good estimate of $\sigma_e=\frac23 0.2=0.133$ for model
III(s=.5) ($\sigma_e=0.130$).

From the table it also evident that interfacial roughness plays an
important role in determining properties.
Consider the 2-cut fields I(s=.5) and III(s=.5).
In Figs.~\ref{modelsab}(i) and \ref{modelsab}(l) it is clear
that both models contain sheet-like pores. The differences are then
due to the interfacial roughness. This is confirmed by comparing the
relative values of $\sigma_u$ for model I in the cases $K=8$ (smooth on
scales below $\lambda_{min}=2\pi/8$) and $K=\infty$
(rough on all scales) - see Appx.~\ref{surfacefd}.
The effect of increasing $K$ from 8 to 32 on the morphology of model
I(s=.5) can be seen by comparing Fig.~\ref{modelsab}(i) \& Fig.~\ref{rough}(a).
In the rough model the sheet-like pores are thinner and a large 
proportion of pore space is distributed in protrusions.
As these protrusions contribute little to the overall conductance
(or strength), this significantly reduces both conductivity and strength.
This also explains why Model III is more conductive (stronger) than
Model I.
The much smaller
effect of roughness on morphology of 1-level cut fields can be seen
by comparing Fig.~\ref{modelsab}(d) \& Fig.~\ref{rough}(b):
the basic inclusion shape is less affected than in the 2-cut case.

Now consider the data for the spherical particulate media in
Table~\ref{qual_table}. The hollow sphere model appears to be more
conductive, or stronger, than the IOS-voids (swiss-cheese) model. 
This is due
to the fact that the former model has an approximately sheet-like
character (Fig.~\ref{IOSAr_8p_2}) in contrast to the node/bond-like
structures~\cite{Isichenko92} apparent in the inverse IOS model.
The IOS model at $p=0.2$ does not
have a sufficient density of spheres to provide a percolative
pathway.

\section{Conclusion}
In this paper we have derived the 3-point statistical correlation
functions for two models of random composite media. The results
were applied in the evaluation of bounds on the effective
conductivity and elastic moduli of each model.
In addition the `exact' effective
conductivity was estimated for the 2-level cut GRF model by direct
simulation.
The models are applicable to physical composites which
remain percolative at very low volume
fractions $p_c<1\%$. These include
solid foams, porous rocks, membranes and sintered hollow glass
spheres. In contrast, previously employed models of
microstructure have percolation thresholds of order 10\%. 

Microstructure was demonstrated to have a strong influence on
the effective properties of composites.
The relative variations amongst the 2-level cut, IOSA and 1-level
cut models were attributed to
three morphological factors; pore-shape, interfacial roughness and
the percolation threshold of the material.
Materials with sheet-like inclusions were
shown to have a significantly greater conductivity/strength
than materials with node/bond-like inclusions. 
By comparing the microstructure parameters of similar composites
with fractally rough and relatively smooth inclusions we found that
interfacial roughness decreased composite conductivity/strength.
The observation was confirmed by directly comparing simulated values of
$\sigma_e$ for Model I(s=.5) (sheet-like and rough) and
Model III(s=.5) (sheet-like and smooth).
The behaviour was physically attributed to the fact that
the protrusions of rough interfaces contribute little to 
effective properties. 

The models discussed here considerably expand the range of systems
to which bounds can be applied.
To facilitate use of these bounds
we have tabulated cross-sections and microstructure parameters
for a number of different variants of each model. Such bounds have two clear
applications. Firstly they can be used to narrow the possible
microstructures of a composite for which properties are known; composite
materials may violate the bounds for a particular model system.
Indeed for certain cases of realistic
media the bounds are mutually exclusive (see
Fig.~\ref{ab10_1}). Secondly, the upper bound is often a very useful estimate of
the actual property. Indeed for model III(s=.5) the upper bound 
provides an excellent estimate of the effective conductivity
over the full range of volume fraction measured.

The simulational data presented here allows comparison of model
properties with those of physical
composites~\cite{Roberts95b,Roberts95d,pRoberts95c}. 
Furthermore the data can be used to assess both predictive theories
for $\sigma_e$, and higher order bounds.  We note that the
4-point correlation functions of the two models considered
here can be calculated, and hence used to evaluate known 4-point
bounds~\cite{Milton82b,Helte95}. 
Finally we remark that the generalization of the IOS model to include
the case of hollow spheres broadens the utility of the model as a 
`bench-mark' theoretical tool; as well as providing a realistic model of
certain composites.
%
%
\appendix
\section{The level cut Gaussian random field}
\label{maxematics}

In this appendix several results are derived which are useful for
calculating the $n$-point correlation function of a material defined
by level cut(s) of a Gaussian random field~\cite{Berk87,Berk91,Teubner91}.
The joint probability density (JPD) of a Gaussian random field is
$P_n(y_1y_2\dots y_n)=( (2\pi)^ n |G|)^{-\frac12}
\exp( -\frac 12 {\mathbf y}^T G^{-1} {\mathbf y})$.
where the elements of $G$ are $g_{ij}=g(r_{ij})=\langle y({\mathbf r}_i)y({\mathbf r}_j)\rangle$
\cite{Wang45}. The function $g$ has the properties $g(0)=1$
and $\lim_{r\to\infty} g(r)\to0$.
By definition we have
$
p_n=\int_{\alpha}^{\beta}\!dy_1\dots\!\int_{\alpha}^{\beta}dy_nP_n(y_1y_2\dots
y_n).$
Note that, in this form, $p_n$ is difficult to evaluate. For example if
$g_{ij}\simeq1$ for all $i,j$ then $|G|\simeq0$ ($n>1$).
It is possible to avoid such problems, and reduce the number of
integrations required, by taking the following approach.

Expanding Eqn.~(\ref{defnpn6}) gives terms of the form,
\begin{equation} \label{Lamb}
\Lambda_n({\mathbf g},{\mathbf a})=
\left\langle \prod_{i=1}^{n} H(y_i-\alpha_i) \right\rangle \end{equation}
where ${\mathbf g}=(g_{12},\dots,g_{(n-1)n})$,
${\mathbf a}=(\alpha_1,\dots,\alpha_n)$ and the $\alpha_i$ are equal to
$\alpha$ or $\beta$.
The analysis which follows relies on an integral representation of
the Heaviside function,
\begin{equation}
H(y-\alpha)=\frac{-1}{2\pi i}\int_C e^{-iw(y-\alpha)}\frac{dw}{w}
\end{equation}         
where the contour $C$ lies along the real axis except near the origin
where it crosses the imaginary axis in the upper half plane.

Now we turn to the evaluation of the terms 
Eqn.~(\ref{Lamb}).
For the case $n=1$ we have $G=g_{11}=1$ so
\begin{equation} \label{lam1}
\Lambda_1=\langle H(y_1-\alpha_1)\rangle=
\frac1{\sqrt{2\pi}}\int_{\alpha_1}^\infty\!\! e^{-\frac12{t^2}}dt.
\end{equation}
Now consider $\Lambda_2$, in this case the matrix $G$ in the JPD is
\begin{equation} G= \left[
\begin{array}{cc} g_{11} & g_{12} \\ g_{21} & g_{22} \end{array} \right]
=\left[ \begin{array}{cc} 1 & g \\ g & 1  \end{array} \right], \end{equation}
with $g=g_{12}=g_{21}$ and $|G|=\sqrt{1-g^2}$.
Using the Heaviside function and interchanging the order of
integration gives,
\begin{eqnarray}
\Lambda_2&=& \nonumber
\frac1{|G|^\frac12(2\pi i)^2}
\int_C \frac{dw_1}{w_1} \int_C\frac{dw_2}{w_2}
e^{i {\mathbf w}^T {\mathbf a} } 
\\ && \times {\int_{-\infty}^{\infty}} {\!\!dy_1}
{\int_{-\infty}^{\infty}} {\!\!dy_2}\;
{e^{-\frac12 {\mathbf y}^T G^{-1} {\mathbf y}-i{\mathbf w}^T{\mathbf y}}} \nonumber
\\&=&
\frac1{(2\pi i)^2}
\int_C \frac{dw_1}{w_1} \int_C\frac{dw_2}{w_2}
e^{i{\mathbf a}^T {\mathbf w} - \frac12w_1^2-w_1w_2g-\frac12w_2^2}.
\nonumber
\end{eqnarray}
In this case we differentiate with respect to $g$
\begin{equation}
\frac{\partial\Lambda_2}{\partial g}=-\frac1{(2\pi i)^2}
{\int_{-\infty}^{\infty}}\!\!dw_1 {\int_{-\infty}^{\infty}} \!\! dw_2\; e^{-\frac12 {\mathbf a}^T G {\mathbf a} + i {\mathbf a}^T {\mathbf w}},
\end{equation}
and perform the integrals with respect to $w_i$. The result is simply
integrated to give $\Lambda_2$ (up to a constant)
\begin{equation} \Lambda_2 = \frac1{(2\pi)} \int_0^g \label{lam2}
\frac{dt}{\sqrt{1-t^2}}
\exp\left({-\frac{\alpha_1^2-2\alpha_1\alpha_2 t +\alpha_2^2}{2(1-t^2)}}\right).
\end{equation}
The derivation of $\Lambda_3$ follows similar lines:
The initial integration over the $y_i$ gives
\begin{equation}
\label{fill1}
\Lambda_3= \frac{-1}{(2\pi i)^3}
\int_C \frac{dw_1}{w_1}
\int_C \frac{dw_2}{w_2}
\int_C \frac{dw_3}{w_3}
e^{-\frac12 {\mathbf w}^T G {\mathbf w}+i {\mathbf w}^T{\mathbf a}}.
\end{equation}
For this case $|G| = 1-g_{12}^2-g_{13}^2-g_{23}^2+2g_{12}g_{13}g_{23}$ and
${\mathbf w}^T G {\mathbf w}=w_1^2+w_2^2+w_3^2+2w_1w_2 g_{12}+2w_1w_3
g_{13}+2w_2w_3 g_{23}$.
Taking the derivative of Eqn.~(\ref{fill1}) with respect to $g_{12}$ gives
\begin{eqnarray}
\frac{\partial \Lambda_3}{\partial g_{12}}&=&
\frac{1}{(2\pi i)^3} \int_C  \frac{dw_3}{w_3} e^{-\frac12 w_3^2
-i\alpha_3 w_3} \\ \nonumber  && \times {\int_{-\infty}^{\infty}}\!\!
dw_1{\int_{-\infty}^{\infty}}\!\!  dw_2 \;
e^{-\frac12 \hat{\mathbf w}^T \hat G
\hat{\mathbf w}+i\hat{\mathbf w}^T{\mathbf u}} \end{eqnarray}
where $\hat{\mathbf w}=(w_1,w_2)$,
${\mathbf u}=(u_1,u_2)=(\alpha_1+ig_{13} w_3,\alpha_2+ig_{23} w_3)$
and
\begin{equation} \hat G= \left[
\begin{array}{cc} 1 & g_{12} \\ g_{12} & 1 \end{array}
\right]. \end{equation}
Performing the standard integrals with respect to $w_1$ and $w_2$ gives,
\begin{eqnarray}
\nonumber
\frac{\partial \Lambda_3}{\partial g_{12}}&=&\frac{1}{2\pi}\frac1{\sqrt{1-g_{12}^2}}\exp
\left({-\frac{\alpha_1^2-2\alpha_1\alpha_2g_{12}+\alpha_2^2}{2(1-g_{12}^2)}}\right)
\\ && \times \left({\frac{-1}{2\pi i}}\right) \int_C\frac{ dw_3}{w_3} e^{-\frac12\nu
w^2+i\kappa w}
\end{eqnarray}
where $\nu=|G|/(1-g_{12}^2)$ and 
\begin{equation} 
\kappa = \alpha_3-
\frac{\alpha_1(g_{13}-g_{12}g_{23})-\alpha_2(g_{23}-g_{12}g_{13})}{1-g_{12}^2}.
\end{equation}
Now the remaining integral can be re-expressed to give
\begin{eqnarray} \nonumber
\frac{\partial \Lambda_3}{\partial g_{12}}&=&\frac{1}{(2\pi)}\frac1{\sqrt{1-g_{12}^2}}\exp
\left({-\frac{\alpha_1^2-2\alpha_1\alpha_2g_{12}+\alpha_2^2}{2(1-g_{12}^2)}}\right)
\\ && \times \frac{1}{\sqrt{2\pi}} \int^\infty_{F_{12}}\!\!
e^{-\frac12 t^2} dt \end{eqnarray}
where $F_{12}=\kappa/\sqrt{\nu}$.
Similar expressions can be derived for $\partial{\Lambda_3}/\partial{g_{12}}$,
and $\partial{\Lambda_3}/\partial{g_{23}}$. These are denoted by
$A_{ij}({\mathbf g},{\mathbf a})=\partial{\Lambda_3}/\partial{g_{ij}}$.
With $k\neq i$ or $j$ we can also write a general expression for $F_{ij}$
\begin{eqnarray} 
F_{ij}&=&\sqrt{\frac{1-g_{ij}^2}{|G|}} \nonumber \\ \nonumber
&&\times \left({\alpha_k-\frac{\alpha_i(g_{ik}-g_{jk}g_{ij})+\alpha_j
(g_{jk}-g_{ik}g_{ij})}{1-g_{ij}^2}}\right).
\end{eqnarray}
The results can be formally integrated to give, up to a constant,
\begin{eqnarray} \Lambda_3({\mathbf g},{\mathbf a}) &=& \nonumber
\int_0^1 \!\! dt \left[ g_{12}A_{12}(t{\mathbf g},{\mathbf a}) \right. \\ && \left.
+g_{13}A_{13}(t{\mathbf g},{\mathbf a}) +g_{23}A_{23}(t{\mathbf g},{\mathbf a})
\label{lam3} \right].
\end{eqnarray}
The results for $\Lambda_i$ are employed in the text to derive the
statistical correlation functions.
\section{Fractal surface dimension}
\label{surfacefd}
Berk \cite{Berk91} has shown that the class of level-cut GRF models
with spectra $\rho(k)\simeq \frac{a}{4\pi} k^{-2\epsilon-3}$ as $k\to\infty$
($0<\epsilon<1$) have field-field correlation functions
$g(r)\simeq 1- b r^{2\epsilon}$ and surface fractal dimension $D_s=3-\epsilon$.
Here $a$ and $b$ are related constants.
In this appendix we show how the finite cut-off wave-number $K$ effects
the roughness (fractal) properties of a GRF interface.
Through a very elegant argument Debye {\it et al} \cite{Debye57} showed that
the surface to volume ratio ($S/V$) of a porous solid was related to
the two point correlation function by
\begin{equation} -4 p_2'(0)=\frac SV . \label{specific} \end{equation}
Now consider $p_2$ for the general 2-level cut Gaussian random field.
The most instructive method of examining $p_2'(0)$ is by
generating an expansion for small $r$.  Thus we write
\begin{equation} p_2(r)=p-\frac1{2\pi}\int_{1-\delta}^{1} \frac{dt}{\sqrt{1-t^2}} f(t), \end{equation}
where $\delta(r)=1-g(r)$, and $f(t)$ is a suitably defined function.
Integrating by parts and retaining leading order terms gives
\begin{equation}\label{p2byparts} p_2(r)\simeq p- {\sqrt{2\delta(r)}}f(1)/{2\pi}, \end{equation}
with $f(1)=\exp(-\alpha^2/2)+\exp(-\beta^2/2)$.
Now if $\delta(r)=O(r^2)$ then the specific surface is well defined and
$p_2'(0)$ can be evaluated.
However for the class of spectra considered by
Berk~\cite{Berk91} $\delta(r)\simeq br^{2\epsilon}$ 
so
\begin{equation} p_2\simeq p-\sqrt{2b}f(1)r^\epsilon/2\pi. \label{p2Ibyparts} \end{equation} Therefore
$p_2'(0)$ and the specific surface ($S/V$) are infinite. 
Bale and Schmidt~\cite{Bale84} have shown that this type 

\begin{figure}[bt!]
{\samepage\columnwidth20.5pc

\hskip 0mm
\centering \epsfxsize=8.0cm\epsfbox{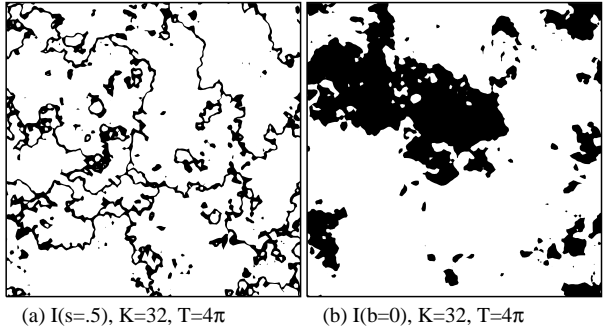}

\vspace{4mm}
\caption{\sl Cross sections of two models based on spectrum I. This
figure shows the roughness of the interface for large $K$ (Compare
Figs.\ \ref{modelsab}(i) \& (d)).
\label{rough}}} \end{figure}

\vspace{-1mm}

\noindent
of singular
behaviour implies a fractal surface. The fractal 
dimension `$D_s$' is given in terms of the correlation function through
the relation $p_2(r)\simeq p-cr^{3-D_s}$ with $c$ some constant.
We infer from Eqn.~(\ref{p2Ibyparts}) that our Model I $(\epsilon=1/2)$
has a fractal surface with $D_s=2.5$.

As discussed in \cite{Roberts95a} it is necessary to introduce a
finite-cutoff wave number $K$ for computational and physical reasons.
We now show how this parameter changes the microstructure.
The wave-number $K$ corresponds to a cut-off wavelength $\lambda_{min}
=2\pi/K$ which specifies the scale of the smallest ``ripples'' on the
surface.  Thus we expect
the surface area to scale as a fractal down to some length scale related
to $\lambda_{min}$.
This can be confirmed mathematically by considering the small $r$
behaviour of $p_2(g_{\mbox{\tiny$K$}}(r))$. 

For arbitrary $r$, $g_{\mbox{\tiny$K$}}(r)$ can be expressed in terms of the moments
of $k$.
Using a Taylor
series expansion for $\sin k r$ in the definition of $g_{\mbox{\tiny$K$}}$
(\ref{gK})
we have
\begin{eqnarray}
g_{\mbox{\tiny$K$}}(r)&=& 1- \frac16 \left({\int_0^K 4\pi \rho_{\mbox{\tiny$K$}}(k){k^4} dk}\right)
r^2+O(K^3r^2) \nonumber \\ \nonumber &\approx& 1-\frac16\langle k^2\rangle r^2,
\end{eqnarray}
where the latter approximation is valid if $r\ll \lambda_{min}$.
Substituting this result into the
expansion for $p_2$ (\ref{p2byparts}) and using relation (\ref{specific}) gives
\cite{Berk87,Teubner91}
\begin{equation} \frac S V =\frac2\pi \sqrt{\frac{\langle k^2\rangle}{3}}\left({
e^{-\frac12{\alpha^2}}+e^{-\frac12{\beta^2}} }\right) \label{grfspecific}. \end{equation}
Thus for $r\ll \lambda_{min}$  the surface is behaving
in a regular manner ($D_s=2$) as anticipated.
Note that for the case $K\to\infty$ and $\epsilon<1$ the moment $\langle k^2\rangle$ diverges
and this approximation does not apply.

To  examine the behaviour for $r > \lambda_{min}$ we can successively
integrate (\ref{gK}) by parts~\cite{Roberts95a} to obtain
\begin{eqnarray}
\label{Iasy} g_{\mbox{\tiny$K$}}(r)&\simeq& \frac{g_\infty(r)}{P} \\ \nonumber && -
\frac{a}{PK^{2\epsilon}}\left({\frac{\cos Kr }{K^2 r^2}+
\frac{3 \sin Kr }{K^3 r^3}+O(K^{-4} r^{-4})}\right). 
\end{eqnarray}

If $Kr >1$ this expansion is asymptotic to $g_{\mbox{\tiny$K$}}$~\cite{Murray00}.
Now in the region $\lambda_{min} \ll r \ll 1$ the algebraic terms in the
expansion are negligible and $g_{\mbox{\tiny$K$}} \simeq g \simeq (1-br^{2\epsilon})/P$
(with $P\simeq1$). 

In summary we have
\begin{equation} p_2(r)\approx
\left\{ \begin{array}{rl}
{ p-\left({\sqrt{ \frac13 \langle k^2 \rangle }\frac{f(1)}{2\pi}}\right)r} &
{0 \leq r \ll \lambda_{min}} \\
{p- \left({\sqrt{ 2b}\frac{f(1)}{2\pi}}\right)r^\epsilon} &
{\lambda_{min} \ll r \ll 1}. \end{array} \right.
\end{equation}
This demonstrates the regular ($D_s=2$) nature of the surface in the
former region, and the fractal behaviour ($D_s=3-\epsilon$) over the spatial
scales in the latter region.
%
\section{Intersection volume of two and three spheres}
\label{ivols}
The intersection volume $V^{(2)}_{I\mu\nu}(d)$ of two spheres of
radii $\mu$ and $\nu$ separated by a distance $d$ is simple to calculate. With
$r_1=\min (\mu,\nu)$ and $r_2=\max(\mu,\nu)$
$V^{(2)}_{I}= 4\pi r_1^3/3$ if $0 \leq d <  r_2-r_1$, 
$V^{(2)}_{I}= 0$ if $r_2+r_1 \leq d < \infty$ and
\begin{equation}
V^{(2)}_{I}=
2\pi (r_1^3+r_2^3)/3 
-\pi ( r_1^2 x_1 + r_2^2 x_2 -\frac13 x_1^3- \frac13 x_2^3 ) 
\end{equation}
if $r_2-r_1  \leq d <  r_2+r_1 $. Here
$x_1=(d^2+r_1^2-r_2^2)/(2d)$ and $x_2=d-x_1$.
 
A compact form of the intersection volume of three spheres of
equal radii ($r=1$) has been derived previously by Powell~\cite{Powell64}.
Several of the key simplifications in the derivation formula are
not possible when the spheres have different radii. However a
less elegant but straight forward result can be determined.
Suppose the spheres have radii $r_A$, $r_B$ and $r_C$ and are distances
$a$, $b$ and $c$ apart and that there exist two unique points
$P$ and $Q$ where the surface of the spheres meet. From Powell~\cite{Powell64}
the intersection volume of the three spheres is equal to twice the following
expression (Powell's Theorem):
\vspace{1mm}

\noindent
\begin{tabular}{rp{8cm}}
&The volume of the tetrahedron PABC
\\ $-$& The volume of the sphere center A enclosed by the faces of the
tetrahedron PABC which meet at A
\\ $-$&  The volume of the sphere center B enclosed by the faces of the
tetrahedron PABC which meet at B
\\ $-$&  The volume of the sphere center C enclosed by the faces of the
tetrahedron PABC which meet at C
\\ $+$&  The intersection volume of the spheres centered at B and C enclosed by
the two faces of the tetrahedron PABC which meet in BC
\\ $+$&  The intersection volume of the spheres centered at C and A enclosed by
the two faces of the tetrahedron PABC which meet in CA
\\ $+$&  The intersection volume of the spheres centered at A and B enclosed by
the two faces of the tetrahedron PABC which meet in AB
\end{tabular}
\vspace{2mm}

\begin{figure}[bt!]
{\samepage\columnwidth20.5pc

\hskip 0mm
\centering \epsfxsize=7.5cm\epsfbox{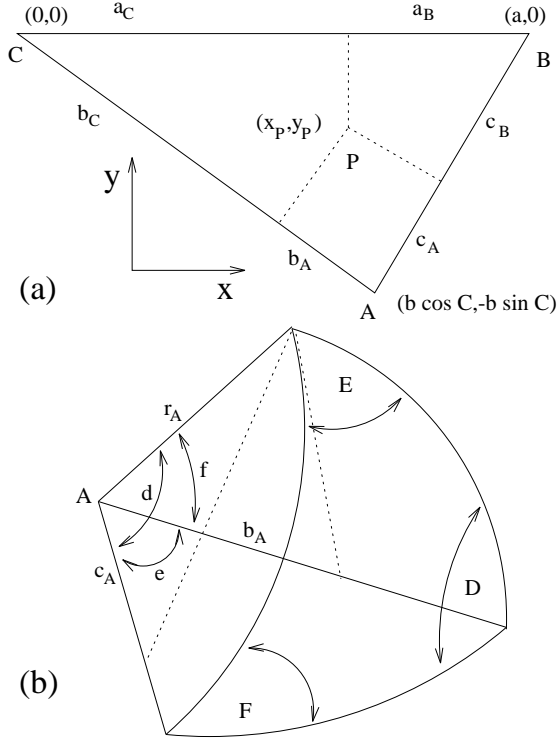}

\vspace{4mm}
\caption{\sl Geometrical elements used in the calculation of the
intersection volume of three spheres of different radii. The axis and
nomenclature are those employed in the text. (a) the tetrahedron base;
(b) the sphere centered at $A$ enclosed by three faces of the
tetrahedron. \label{fig3i1}}} \end{figure}

\vspace{-1mm}

\noindent
The cases where there is no unique point of intersection between the
spheres is discussed below.  We first define a coordinate system
with origin at the center of sphere $C$ as drawn in Fig.~\ref{fig3i1}(a).
By solving the equations of the
three spheres simultaneously it is simple to show that
\begin{eqnarray}
x_P&=&{{{a^2} - {{r_{B}}^2} + {{r_{{\rm C}}}^2}}\over {2\,a}} \\
y_P&=&\frac{-b^2+r_A^2-r_C^2 + 2b \cos C x_P }{2b \sin C} \\
z_P&=&\sqrt{r_C^2-x_P^2-y_P^2}.
\end{eqnarray}
It is also necessary to know the distances $a_B,a_C\dots$ given in
Fig.~\ref{fig3i1}(a).
We have $a_C=(a^2+r_C^2-r_B^2)/(2 a)$, $b_A=(b^2+r_A^2-r_C^2)/(2b)$
$c_B=(c^2+r_B^2-r_A^2)/(2c)$, $a_B=a-a_C$, $b_A=b-b_C$ and $c_A=c-c_B$.

The volume of the tetrahedron is $V_T=\frac16 ab \sin C z_P$.
The solid angle $\phi_A$ of the tetragonal wedge at A (see 
Fig.~\ref{fig3i1}(b)) can be calculated by using the fact that
$\phi_A= (E+F+D-\pi)$ and 
\begin{equation}\cos D=\frac{ \cos d - \cos e \cos f}{ \sin e \sin f } \end{equation}
(similarly for $\cos E$ and $\cos F$). This gives

\begin{figure}[bt!]
{\samepage\columnwidth20.5pc

\hskip 0mm
\centering \epsfxsize=7.5cm\epsfbox{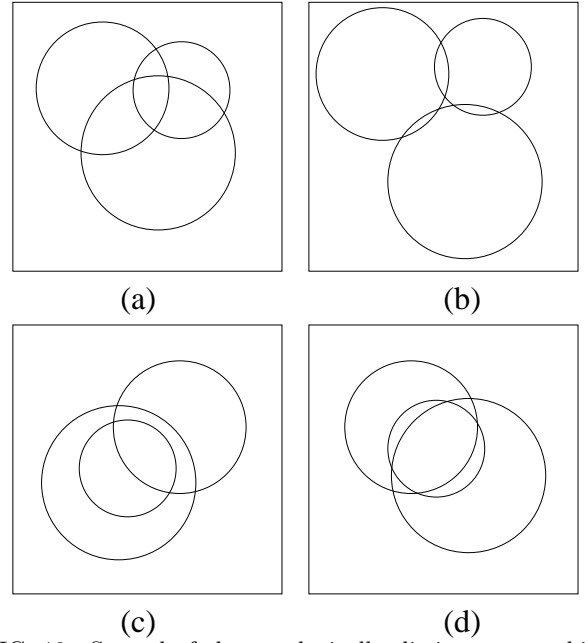}

\vspace{1mm}
\caption{\sl Several of the topologically
distinct cases which arise in the calculation of the intersection
volume of three spheres of different radii. \label{casesiosa}}}
\end{figure}

\vspace{-2mm}

\noindent
\begin{eqnarray} \nonumber
\phi_A&=&\cos^{-1} \left({\frac{c_A -\cos A b_A}
{\sin A \sqrt{r_A^2-b_A^2}}}\right)\\
&&+\cos^{-1} \left({\frac{r_A^2 \cos A- b_A c_A}{ \sqrt{r_A^2-b_A^2}
\sqrt{r_A^2-c_A^2}}}\right) \\
&&+\cos^{-1} \left({\frac{b_A -\cos A c_A}{\sin A \sqrt{r_A^2-c_A^2}}}\right)-\pi.
\end{eqnarray}
Similar results are obtained for the solid angles $\phi_{B,C}$.

It is critical to know whether the point $x_P,y_P$ lies inside or outside
each of the faces of the triangle. This can be done by defining the
variables
\begin{eqnarray}
s_A&=&{\rm sgn}(-y_P) \\
s_B&=&{\rm sgn}(\cos C y_P+ \sin C x_P) \\
s_C&=&{\rm sgn}(\cos B y_P- \sin B x_P +a \sin B).
\end{eqnarray}
Then for example $s_A=\pm1$ as the point $(x_P,y_P)$ is inside or
outside face $a$ of the triangle $ABC$.
In the case $r_A=r_B=r_C=1$ (Powell~\cite{Powell64}) 
we have $x_P=a/2$ and $y_P=-c\cos A /2\sin C$ so that
$s_A={\rm sgn}( \cos A)$, $s_B={\rm sgn}( \cos B)$ and $s_C={\rm sgn}( \cos C)$ as
they should.

The wedge angle associated with the intersection volume of spheres
B \& C is,
\begin{equation} \theta_A=\cos^{-1}\left({ s_A \frac {\sqrt{r_B^2-a_B^2-z_P^2}}
{\sqrt{r_B^2-a_B^2}}}\right).  \end{equation}
Similarly for the angles $\theta_B$ \& $\theta_C$.

Now the volume of a tetragonal wedge of solid angle $\phi$ is $r^3 \phi/3$
and the intersection volume of spheres enclosed in a wedge of
angle $\theta$ is $\theta V^{(2)}_I/ 2\pi$. Therefore, by Powell's theorem,
\begin{eqnarray} V_{Ixyz}^{(3)}(a,b,c)&=&2 V_T - \frac23 x^3 \phi_A
- \frac23 y^3 \phi_B- \frac23 z^3 \phi_C \label{vithree} \\ &&+
 \frac{\theta_A }{\pi} V^{(2)}_{Iyz}(a)
+\frac{\theta_B }{\pi} V^{(2)}_{Ixz}(b)
+\frac{\theta_C }{\pi} V^{(2)}_{Ixy}(c). \nonumber
\end{eqnarray}
Here $x=r_A$, $y=r_B$ \& $z=r_C$.  This formula is equivalent to Powell's
result~\cite{Powell64} in the case $x=y=z=1$. 

Several other cases arise if the point $P$ does not exist. Some of these
are illustrated in Fig.~\ref{casesiosa}. Either two (or more) of the spheres are
disconnected (not illustrated), they are connected but
$Va^{(3)}_I=0$ (b) or the intersection volume is given by that of
two of the spheres (c) or some other formula (d).

\vspace{-5mm}

\section{Derivation of $\zeta_1|_{p=0}$ (IOSA)}
\label{confzeta}

\vspace{-3mm}
It is possible to develop an independent check on the calculation of
$\zeta_1$ for the IOSA model by direct calculation of $\sigma_e$. 
Using the framework of Reynolds
and Hough~\cite{Reynolds57} gives
\begin{equation} \label{Rey1}
\sigma_e=\sigma_2+(\sigma_1-\sigma_2) p f_1 \end{equation}
where $f_1=\bar E_1 / \bar E$. Here $\bar E_1$ is the average of the
field throughout phase 1 and $\bar E$ is the applied field.
While the above formula is exact it is only possible to evaluate
$E_1$ approximately. In the low concentration
regime ($p\ll1$) $E_1$ is the field within a
hollow sphere (conductivity $\sigma_1$) embedded in an infinite
medium (conductivity $\sigma_2$) subject to an applied
field $\bar E$. To determine this field we
consider a more general problem where the conductivities of the
innermost spherical region ($0\leq r< r_0$), the annulus ($r_0 \leq r < r_1$)
and the enclosing medium ($r_1 \leq r < \infty$) are $\sigma_a$,
$\sigma_b$ and $\sigma_c$ respectively.
The potential of the field satisfies Laplace's equation and 
charge conservation boundary conditions at phase boundaries.
Using standard techniques it is possible to
show that, in each region, the potential has the form
$\phi_d=( A_d r + B_d r^{-2}) \cos \theta$ with $d=a,b$ or $c$.
Applying the appropriate boundary conditions on each of the faces
of the hollow sphere gives,
\begin{eqnarray} \nonumber
A_a&=&-9 s^3 \bar E/H;\;\; B_a=0;\;\; A_b =  -3 s^3 \bar E (2+x)/H;
\\ \nonumber
B_b&=& -3 r_1^3 \bar E (1-x)/H;\;\; A_c = -\bar E; \\
B_c&=&-r_1^3 \bar E [(1-x)(1+2y)+s^3 (2+x) (1-y)]/H.
\nonumber
\end{eqnarray}
where $H=2(1-x)(1-y)+s^3(2+x)(2+y)$, $x=\sigma_a/\sigma_b$,
$y=\sigma_b/\sigma_c$ and $s=r_1/r_0$.
For the desired value of $f_1$, $\sigma_a=\sigma_c=\sigma_2$ and
$\sigma_b=\sigma_1$. Considering volume averages of the field lead to 
\begin{equation} 
(f_1)_z=-\frac{A_b}{\bar E}=\frac{3 s^3 \sigma_2 ( 2\sigma_1+\sigma_2)}
{ s^3 ( 2\sigma_1+\sigma_2)( 2\sigma_2+\sigma_1)- 2 (\delta\sigma)^2},
\nonumber
\end{equation}
where $\delta\sigma=\sigma_1-\sigma_2$ and $(f_1)_x=(f_1)_y=0$.
Now expanding Eqn.~(\ref{Rey1}) in powers of $\delta\sigma$ gives,
\begin{equation} \sigma_e\simeq \sigma_2 + (\delta\sigma) p - \frac {1}{3\sigma_2} (\delta\sigma)^2 p
+ \frac { 2 + s^3}{9 s^3 \sigma_2^2} (\delta\sigma)^3 p.  \end{equation}

\begin{figure}
{\samepage\columnwidth20.5pc

\hskip 0mm
\centering \epsfxsize=7.5cm\epsfbox{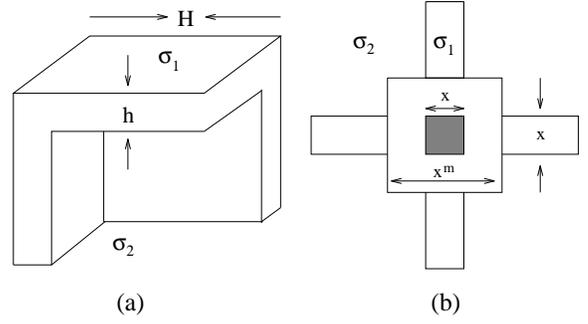}

\vspace{2mm}
\caption{\sl Periodic cellular models:
(a) Sheet-like cell;
(b) Bond/node-like cell.
\label{unitcell}}}
\end{figure}

\vspace{-2mm}

\noindent
Similarly Brown's formula~\cite{Brown55} to the same order gives,
\begin{equation} \sigma_e\simeq \sigma_2 + (\delta\sigma) p - \frac {1}{3\sigma_2} (\delta\sigma)^2 p
+ \frac { 1 + 2\zeta_1}{9 \sigma_2^2} (\delta\sigma)^3 p.  \end{equation}
Equating similar terms leads to $\zeta_1|_{p=0}=s^{-3}=r_0^3/r_1^3$.
Points representing this result are plotted in Fig.~\ref{z_ab_sum} and confirm
prior calculations of $\zeta_1$. It should also be possible to calculate
the first order correction, ${\partial \zeta_1}/{\partial p}|_{p=0}$,
by calculating
$\sigma_e$ to $O(p^2)$~\cite{Thovert90,Jeffrey73}. Since $\zeta_1$
is observed to have a linear behaviour over a wide range of
$p$~\cite{Thovert90} (see Fig.~\ref{z_ab_sum}) this would provide a good
estimate of $\zeta_1$. Also note that $\eta_1|_{p=0}$ can be derived
using similar methods.

\vspace{-5mm}

\section{Periodic cell models}
\label{cellfoam}

\vspace{-3mm}

To explicitly demonstrate the effect of pore shape on effective
conductivity we estimate $\sigma_e$ for several periodic networks
exhibiting sheet-like, grid-like and node/bond-like cells.
Consider a structure comprised of periodic repetitions of the
unit cell shown in Fig.~\ref{unitcell}(a).
Defining $x=h/H$ the volume
fractions of each phase are given by $p=1-(1-x)^3 \simeq 3x$ and
$q=(1-x)^3   \simeq 1-3x$.
Consider the behaviour of the model if $\sigma_1 \gg \sigma_2$. In this case
most of the current would flow through the solid faces of the cell
which are aligned in the direction of current flow. The volume
fraction of these elements of the cell is $p_1=2x-x^2$.
The remaining current would pass through a layer
of phase 1 (volume fraction $p_2=x(1-x)^2$) and the cell core of phase 2
(volume fraction $q$).
Treating each of these mechanisms as conductors in parallel we have
$\sigma_e=p_1 \sigma_1 + (p_2+q) \sigma^*$, where $\sigma^*$ is
conductivity of the central leakage pathways. Assuming each of the
elements of these pathways act as conductors in series
gives $\sigma^*=(p_2+q)(p_2\sigma_1^{-1}+ q\sigma_2^{-1})^{-1}$.
This leads to
\begin{eqnarray}
\nonumber
\sigma_e&=&(2x-x^2)\sigma_1+\frac{(1-x)^2\sigma_2}
{x(\sigma_2/\sigma_1)+(1-x)} \\ \nonumber &\simeq&
\sigma_2+\frac23p\sigma_1-\frac13 p\sigma_2\left({1+\frac{\sigma_2}{\sigma_1}}\right)
\end{eqnarray}
where the approximation holds for $p\ll1$.
Finally, $\sigma_e\simeq \frac23 \sigma_1 p$ in the case $\sigma_2=0$.

In a similar way a `toy' model can be defined to qualitatively
demonstrate the effect that necks/throat have on the effective
conductivity.
A cross section of the unit cell of a {\it node/bond}
model is shown in Fig.~\ref{unitcell}(b). The central cube has side length
$x^m$ and the six arms have a square cross section of side length $x$.
Taking the cell to have unit width we have: $p=x^{3m}+3(1-x^m)x^2$ ($m\leq1$) and
$q=1-p$.  If $\sigma_2=0$ then most of the current will flow through the
bonds parallel to the direction of the applied field. Therefore,
$\sigma_e \simeq \sigma_1 x^2 $. In the case $m=1$ a uniform grid results
and $\sigma_e \simeq \frac13 p \sigma_1$ to leading order in $p$.
For $m=1/3$ a node/bond geometry results and $\sigma_e \simeq p^2 \sigma_1$.

\vspace{-5mm}


\end{multicols}

\clearpage

%

%

%

%
%
%
%
%

\clearpage
\widetext
%



\end{document}